\documentclass{amsart}

\usepackage[T1]{fontenc}
\usepackage{authblk}
\usepackage[foot]{amsaddr}
\usepackage{amssymb}
\usepackage{xcolor}
\usepackage{amsmath}
\usepackage{mathrsfs}
\usepackage{algorithm2e}
\usepackage{subfigure}
\usepackage{algpseudocode}
\usepackage{graphicx}
\usepackage[labelfont=bf]{caption}
\newcommand{\comment}[1]{}

\DeclareMathOperator*{\argmin}{arg\,min}
\DeclareMathOperator{\Var}{Var}

\usepackage[margin=0.85in]{geometry}

\title{Optimizing Shift Selection in Multilevel Monte Carlo for Disconnected Diagrams in Lattice QCD}

\author{Travis Whyte$^1$, Andreas Stathopoulos$^1$}
\address{$^1$Department of Computer Science, William \& Mary, Williamsburg VA}

\author{Eloy Romero$^2$}
\address{$^2$Thomas Jefferson National Accelerator Facility, Newport News VA}

\author{Kostas Orginos$^{2,3}$}
\address{$^3$Department of Physics, William \& Mary, Williamsburg VA}

\date{}

\begin{document}

\maketitle

\begin{abstract}
The calculation of disconnected diagram contributions to physical signals is a computationally expensive task in Lattice QCD. To extract the physical signal, the trace of the inverse Lattice Dirac operator, a large sparse matrix, must be stochastically estimated. Because the variance of the stochastic estimator is typically large, variance reduction techniques must be employed. Multilevel Monte Carlo (MLMC) methods reduce the variance of the trace estimator by utilizing a telescoping sequence of estimators. Frequency Splitting is one such method that uses a sequence of inverses of shifted operators to estimate the trace of the inverse lattice Dirac operator, however there is no a priori way to select the shifts that minimize the cost of the multilevel trace estimation. In this article, we present a sampling and interpolation scheme that is able to predict the variances associated with Frequency Splitting under displacements of the underlying space time lattice. The interpolation scheme is able to predict the variances to high accuracy and therefore choose shifts that correspond to an approximate minimum of the cost for the trace estimation. We show that Frequency Splitting with the chosen shifts displays significant speedups over multigrid deflation, and that these shifts can be used for multiple configurations within the same ensemble with no penalty to performance.  
\end{abstract}


\section{Introduction}

Lattice Quantum Chromodynamics (LQCD) is the foremost theoretical tool for estimating physical properties of hadrons. In the extraction of physical observables, it is often required to calculate the contribution of disconnected diagrams, which involves calculating the trace of the inverse of the lattice Dirac operator, $D$. Often, it is not only the trace of $D^{-1}$ that is desired, but the trace of $\Gamma D^{-1}$, where $\Gamma$ is a unitary rotation of $D^{-1}$ corresponding to a desired physical observable. In addition, our motivation comes from the calculation of the disconnected contribution to the flavor separated Generalized Parton functions \cite{PhysRevLett.126.102003,Gambhir:2016EM}, which requires the calculation of $\textbf{Tr}(\Gamma \Omega_p D^{-1})$ where $\Omega_p = W \Pi_p$.
$\Pi_p$ is a permutation matrix corresponding to
a $[0,0,p,0]$-displacement that we assume is in the $z$ lattice direction without loss of generality, and $W$ is the Wilson line, a product of gauge links in the $z$-direction.
The lattice Dirac operator is a large sparse matrix, so explicitly calculating the trace of the inverse is computationally infeasible. The only resort is then to estimate the trace stochastically. Generally, for a non singular matrix $A \in \mathbb{C}^{N \times N}$, the Hutchinson method \cite{doi:10.1080/03610918908812806} estimates the trace of the inverse as 
$\textbf{Tr}(A^{-1}) = \mathbb{E}[z^{\dag}A^{-1}z] = \mathbb{E}(t(A^{-1}))$, i.e., with the estimator
    \begin{equation}
        t(A^{-1}) =  \frac{1}{s}\sum_{j=1}^{N_s} z_j^{\dag}A^{-1}z_j,
        \label{test}
    \end{equation}
where $z_j$ are $N_s$ random vectors whose elements are drawn from $\{1,-1,i,-i\}$ with equal probability. The variance of the estimator is given by
    \begin{equation}
        \Var[t(A^{-1})] = 2(|| A^{-1} ||^{2}_F - \sum_{i=1}^N |(A^{-1})_{ii}|^2),
        \label{test_var}
    \end{equation}
which depends on the magnitude of the off-diagonal elements of $A^{-1}$. In lattice QCD the ill-conditioning of $D$ and thus the above variance grow with the lattice size. It is therefore necessary to develop variance reduction methods. 

Probing \cite{probing} is one variance reduction technique, wherein a coloring of the graph of $A$ results in a set of orthogonal vectors that remove specific nonzero elements of $A^{-1}$. Hierarchical probing \cite{HP,ExtHP} is an extension of probing that allows for reuse of the quadratures in the stochastic trace estimation by identifying nested colorings at increasing lattice distances.  Deflation \cite{Defl,BARAL201964} computes the singular pairs corresponding to the smallest singular values of $A$, which contribute the most to the variance. It has been shown in \cite{Defl} that deflation and probing act complementary to one another, with probing removing the heaviest elements in short lattice distances and deflation removing the contribution of elements at long distances. While effective, deflation poses a problem for LQCD practitioners. 
As the lattice volume grows, the density of the low lying singular space of $D$ increases, and therefore more singular triplets must be deflated out to achieve similar reductions in variance. Therefore, the resulting memory and computational costs to compute and apply deflation become prohibitive.
Multigrid deflation \cite{ROMERO2020109356} was developed as a method to mitigate these scaling costs of deflation. In multigrid deflation, the singular triplets of a coarsened lattice Dirac operator are calculated, rather than those of $D$. Due to the reduced size of the coarsened operator, many more singular triplets can be calculated in comparison to that of $D$. 

Multilevel Monte Carlo methods reduce the variance of the estimator by utilizing a telescoping sequence of estimators \cite{doi:10.1287/opre.1070.0496} with each term in the sequence corresponding to a level $l$. Given the sequence $X_0,X_1,...,X_L$, the unbiased estimator of $X_L$ can be written as
    \begin{equation}
        \mathbb{E}[X_L] = \mathbb{E}[X_0] + \sum_{l=1}^L\mathbb{E}[X_l - X_{l-1}].
    \end{equation}
The variance of the multilevel estimator will then be lower than that of the single level method due to the correlation between $X_l$ and $X_{l-1}$, and will have a lower computational cost than that of the single level method. Recent examples of multilevel trace estimation can be found in \cite{HALLMAN2022125}, where a sequence of Chebyshev polynomials were used to estimate the trace of $f(A)$, and in \cite{multigrid_trace_est}, where $\textbf{Tr}(A^{-1})$ was estimated using a hierarchy of coarsened operators from a multigrid construction. 

A third multilevel method, and the one that is the focus of our study, is Frequency Splitting (FS) \cite{FreqSplit}. FS splits the low and high frequency modes of the inverse lattice Dirac operator by creating a telescoping series of inverses of shifted operators. FS then avoids the computation and storage of singular vectors, which is known to scale as $O(V^3)$ or higher for a lattice of volume V \cite{L_scher_2007}, making it a good candidate for variance reduction in the exascale regime. We also expect the variance associated with the FS method to be smaller than that of deflation when deflating with a number of singular pairs that is practical to compute, due to the rapid decay of the off-diagonal elements of $D^{-1}$ at large quark mass. However, FS is not without its own complications. There is no \textit{a priori} way to predict the optimal shifts that give the minimum multilevel cost. 

The scalability limitations of deflation with respect to lattice volume and the clear advantage of a shifting strategy can be demonstrated by examining the approximate decay coefficient of the inverse. The magnitude of the elements of the inverse Dirac operator decay approximately according to $e^{-\rho|x-y|}$ for large source and sink separation, where $\rho$ is proportional to the pion mass and thus is dependent on the shift $\sigma$ \cite{BOWLER1983137}.
Therefore, the variance of the trace estimator in (\ref{test_var}) decreases as the lattice Dirac operator is shifted. We illustrate this in the following experiment.

\begin{figure}[th]
\centering
\includegraphics[width=1.0\textwidth,scale=0.5]{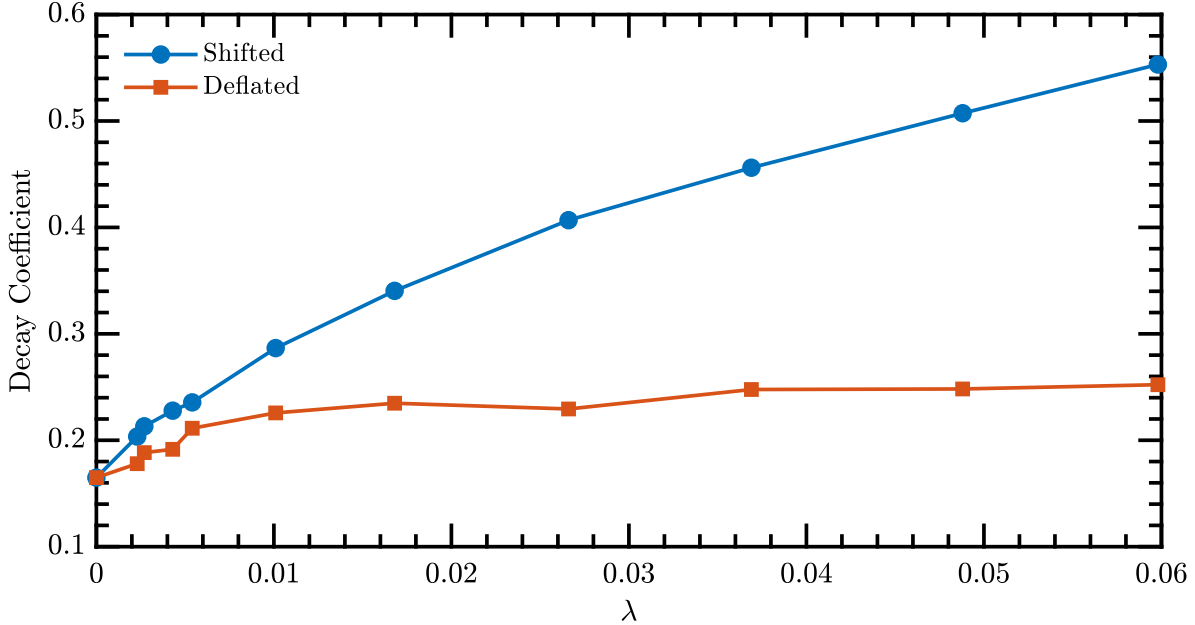}
\caption{The decay coefficient extracted from $(D+\lambda I)^{-1}$ and $(I-P(\lambda))D^{-1}$, where $P(\lambda)$ is the projector to the low singular space that includes all singular triplets up to $\lambda$.}
\label{rho_v_lambda}
\end{figure}

The inverse was sampled by randomly choosing a lattice point and solving the set of linear equations for each color spin index on a lattice of size $32^3 \times 64$ at physical quark mass. The lowest 200 singular values $\lambda_i$ and their singular vectors were calculated and used to deflate the inverse using $i = \{0, 1, 2, 4, 8, 16, 32, 64, 100, 150, 200\}$ singular triplets. We then use the corresponding value of $\lambda_i$ to shift the lattice Dirac operator. In both cases, the value of $\rho$ is extracted from an exponential fit of the propagator as a function of the Euclidean distance between source and sink. It can be readily seen in Figure \ref{rho_v_lambda} that the decay coefficient of the inverse rapidly increases, and thus the variance decreases when shifting by $\lambda_i$. For deflation, the decay coefficient plateaus very early on. It would take deflating many thousands of singular pairs to achieve the same rate of decay and variance reduction as shifting due to the density of the low singular space. This observation further motivates our use of FS and the need to obtain near optimal shifts. 

The rest of the paper follows the following structure: Section \ref{sec:back} discusses probing for displacements of the lattice, FS, and multilevel Monte Carlo in the context of FS. Section \ref{sec:methods} discusses our sampling and interpolation technique which chooses near optimal shifts for the multilevel trace estimation. Section \ref{sec:results} gives experimental results using the chosen shifts to compare to multigrid deflation, and Section \ref{sec:conc} gives our conclusions and resulting open avenues of study.

\section{Background}
\label{sec:back}

\subsection{Probing}
\label{sec:probing}
Probing \cite{probing} is a method that has been used in many applications, but can be used in trace estimation to reduce the variance of the estimator. 
Many matrices, such as the lattice Dirac operator $D$, exhibit decay of the off-diagonal elements of the inverse with respect to the distance of their corresponding points in the graph of the matrix or in the lattice. 
Classical Probing  eliminates elements that correspond to distances up to $k$ by computing a distance-$k$ coloring of the graph of $A$
(which is the same  as the distance-$1$ coloring of $A^k$). Since coloring optimality is not important, the task is performed with a greedy coloring algorithm. 
The resulting $c$ colors are then used to construct a set of orthogonal probing vectors $z_j$, $j = 1,2,...,c$,
    \begin{equation}
        z_j(i) = \begin{cases}
                    1 \quad & \text{if color($i$) $= j$} \\
                    0 \quad & \text{otherwise} \\
                    \end{cases}.
    \label{probing_vectors}
    \end{equation}
Note that these vectors calculate exactly $\textbf{Tr}(A^k)=\sum_{j=1}^c z_j^T A^k z_j$. When used for $\textbf{Tr}(A^{-1})$ they have the desired effect to eliminate from the variance in (\ref{test_var}) all elements corresponding to distances up to $k$.

Recently, Classical Probing has been extended when an estimation of the trace of $\Pi_p A^{-1}$ is desired, where $\Pi_p$ is a permutation matrix that corresponds to a displacement $p$ in the underlying four dimensional space-time lattice \cite{switzer2021probing}. In our LQCD application, we are interested in displacements in the $z$ dimension of the lattice. 
The coloring can then be performed on the symmetric part of $\Pi_pA^k$, given by $\Pi_pA^k + (\Pi_pA^k)^T$. For a node $x=[x_1,...,x_4]$ in the lattice, this corresponds to a distance-$k$ coloring not of the neighborhood centered at $x$ but of the neighborhoods centered at $x^+ = [x_1 ,...,x_4+p]$ and $x^-=[x_1,...,x_4-p]$. 
The probing vectors produced with the resulting colors and (\ref{probing_vectors}) are then used in the trace estimation of $\Pi_pA^{-1}$. They have the effect of removing the contribution of neighboring nodes up to distance $k$ from any node displaced by $p$ with the same color. These will be the heaviest off-diagonal elements of $\Pi_p A^{-1}$. Finally, although the probing vectors $z_j$ are deterministic, we can combine them with a set of $s$ random noise vectors, which is equivalent to taking $s$ steps of Hutchinson on the matrix $\Pi_pA^{-1}\odot ZZ^T$, where the columns of $Z$ are the probing vectors and $\odot$ is the element-wise product.

\subsection{Frequency Splitting}
Given a set of $L$ real shifts of increasing magnitude, $\sigma$, $D^{-1}$ can be written as a telescoping sum of inverses separated by the shifts
    \begin{equation}
        D^{-1} = \sum_{l=0}^{L-1} \left( (D+\sigma_l I)^{-1} - (D+\sigma_{l+1}I)^{-1} \right) + (D+\sigma_L I)^{-1}
        \label{tele_sum}
    \end{equation}
where $0 = \sigma_0 < \sigma_1 < ... < \sigma_L$. FS then makes use of the identity 
    \begin{equation}
        A^{-1}(A - B)B^{-1} = B^{-1}-A^{-1}
    \end{equation}
for $A$ and $B$ square, non singular matrices. In LQCD, this identity was first introduced as the ``One End Trick" and was used in the stochastic estimation of the trace for the twisted mass discretization of fermions \cite{BOUCAUD2008695}. In FS, the inverses differ only by a shift, so Equation (\ref{tele_sum}) can be written as
    \begin{equation}
        D^{-1} = \sum_{l=0}^{L-1} (\sigma_{l+1} - \sigma_{l})(D+\sigma_l  I)^{-1}(D+\sigma_{l+1}I)^{-1} + (D+\sigma_L I)^{-1}.
        \label{tele_sum_prod}
    \end{equation}
This can be generalized to the product $\Gamma D^{-1}$ through left multiplication with any matrix $\Gamma$. Then, taking the trace yields
\begin{center}
    \begin{equation}
            \label{gamma_tele_sum_prod}
        \begin{split}
        \textbf{Tr}(\Gamma D^{-1}) & = \sum_{l=0}^{L-1} (\sigma_{l+1} - \sigma_{l})\textbf{Tr}\left(\Gamma (D+\sigma_lI)^{-1}(D+\sigma_{l+1}I)^{-1}\right) \\
        & + \textbf{Tr}(\Gamma (D+\sigma_LI)^{-1}).
            \end{split}
    \end{equation}
\end{center}
In Ref. \cite{FreqSplit}, the trace estimator of any term within the summation is referred to as the ``standard random-noise estimator". It was further shown that the terms within the summation can be rewritten using the commutation of the inverses and the cyclic property of the trace, yielding
    \begin{align}
    \begin{split}
    \textbf{Tr}(\Gamma (D+\sigma_l I)^{-1}(D+\sigma_{l+1}I)^{-1})
    & = \textbf{Tr}((D+\sigma_l I)^{-1} \Gamma (D+\sigma_{l+1}I)^{-1}).
    \end{split}
    \label{cyclic_commute}
    \end{align}
The trace estimator of the right hand side is known as the ``split-even estimator" and was shown to have lower variance than that of the standard random-noise estimator. While Equation (\ref{cyclic_commute}) shows the equality of the traces, $\Gamma (D+\sigma_l I)^{-1} (D+\sigma_{l+1} I)^{-1} \neq (D+\sigma_l I)^{-1} \Gamma (D+\sigma_{l+1} I)^{-1}$. The insertion of $\Gamma$ in between the two inverses results in a change of the singular spectra, which accounts for the reduction in the variance of the split-even estimator. Equation (\ref{gamma_tele_sum_prod}) can be further generalized using similar means to our target matrix $\Gamma \Omega_p D^{-1}$ for any combination of $\Gamma$ and $\Omega_p$,
    \begin{equation}
    \label{tele_sum_prod_disp0}
        \Gamma \Pi_p D^{-1} = \sum_{l=0}^{L-1} (\sigma_{l+1} - \sigma_{l})\Gamma \Pi_p(D+\sigma_l I)^{-1}(D+\sigma_{l+1}I)^{-1} + \Gamma \Pi_p(D+\sigma_L I)^{-1}.
    \end{equation}
Taking the trace, using its cyclic property, and the fact that inverses of shifted matrices commute, Equation (\ref{tele_sum_prod_disp0}) can be written as
    \begin{equation}
            \label{tele_sum_prod_disp}
        \begin{split}
        \textbf{Tr}(\Gamma \Omega_p D^{-1}) & = \sum_{l=0}^{L-1} (\sigma_{l+1} - \sigma_{l})\textbf{Tr}\left( (D+\sigma_l  I)^{-1}\Gamma \Omega_p (D+\sigma_{l+1}I)^{-1}\right) \\
        & + \textbf{Tr}(\Gamma \Omega_p (D+\sigma_LI)^{-1}).
            \end{split}
    \end{equation}
The trace of each term in Equation (\ref{tele_sum_prod_disp}) can then be estimated independently. The variances of these trace estimators given by Equation (\ref{test_var}) and can be defined as a function, $V_l$ and $V_L$, of $\Gamma$, $\Omega_p$, and shifts $\alpha$ and $\beta$, for $l=0,\ldots,L-1$
\begin{equation}
    V_l(\alpha,\beta, \Gamma, \Omega_p) =  Var[t((\alpha-\beta)(D+\alpha I)^{-1}\Gamma \Omega_p (D+\beta I)^{-1})]
    \label{phi_prod}
\end{equation}
and for the last term as a function of one shift $\tau$,
\begin{equation}
    V_L(\tau, \Gamma, \Omega_p) = Var[t(\Gamma \Omega_p (D+\tau I)^{-1})].
    \label{phi_single}
\end{equation}
When $\Gamma$ and $\Omega_p$ are implied in the context, we simplify the notation as $V_l(\alpha, \beta)$ and $V_L(\tau)$. When all parameters are implied we refer to the variances as $V_l$ and $V_L$.

We remark that when $\Gamma = \Omega_p = I$, any variance reduction achieved is due to the correlation between $(D+\sigma_l I)^{-1}(D+\sigma_l I)^{-1}$ and $(D+\sigma_{l+1}I)^{-1}(D+\sigma_{l+1}I)^{-1}$ due to the equality of the One End Trick. As the shift is increased, the inverses become less correlated and therefore many shifts would need to be taken in order to effectively reduce the variance. It is also prudent to mention here that deflation is not a viable method of variance reduction to use in conjunction with FS. Due to the complex, non symmetric nature of $D$, the insertion of $\Gamma \Omega_p$ in between the two inverses means that the terms within the summation will have different spectra for every combination of $\Gamma$ and $\Omega_p$. The effect of shifting $D$ further changes the singular spectrum. Variance reduction with deflation would then require calculating singular pairs for every product term associated with an individual combination of $\Gamma$ and $\Omega_p$ for every shift. As an example, if the trace of $\Gamma \Omega_p D^{-1}$ is being estimated for 16 different combinations of the $\Gamma$ matrices, 9 different displacements of the lattice, and six shifts, there are 864 split even estimators that require the calculation of singular vectors in order to deflate. Such a calculation is too expensive. FS poses an additional complication: because the relation of the singular values of $D$ to the shift $\sigma_l$ are not known, the variance of the individual terms in Equation (\ref{tele_sum_prod_disp}) cannot be predicted analytically. 
\comment{In Section \ref{sec:methods}\ref{sec:methods}, we describe our sampling and interpolation method that chooses near optimal shifts for minimizing the cost of the trace estimation of a given (\Gamma, \Omega_p)(\Gamma, \Omega_p) pair.}

\subsection{Multilevel Monte Carlo}
\label{subsec:mlmc}
In this section, we introduce the concepts of optimizing the multilevel Monte Carlo in the context of our FS application. An open question with regard to FS was whether it reduced the total computational work compared to estimating the trace of $\Gamma \Omega_p D^{-1}$ with other variance reduction techniques. To answer this question, we first follow the work of Giles for multilevel Monte Carlo  \cite{doi:10.1287/opre.1070.0496} to adjust how accurately we must solve each level so that the cost to estimate Equation (\ref{tele_sum_prod_disp}) is minimized for a particular $\Gamma$ and $\Omega_p$. 

For a single level method, the total computational cost of estimating \textbf{Tr}($\Gamma \Omega_p D^{-1}$) to a target variance $\epsilon^2$ is given by
    \begin{equation}
        \epsilon^{-2}CV,   \label{sl_cost}
    \end{equation}
where $C$ is the solver cost of the linear equations and $V$ is the estimator variance,
$V=\Var[t(\Gamma \Omega_p D^{-1})]$.
For our multilevel FS application, the total computational cost of estimating Equation (\ref{tele_sum_prod_disp}) is given by
    \begin{equation}
        C_{FS} = \epsilon^{-2} \bigg( \sum_{l=0}^{L} \sqrt{C_l V_l} \bigg) ^2,
    \label{ml_cost_shifts}
    \end{equation}
where $\epsilon^{-2}$ is similarly defined, and $C_l$ and $V_l$ are respectively the solver cost and estimator variance for a given set of shifts, $\Gamma$, and $\Omega_p$ at level $l$.
Specifically, $V_l = V_l(\sigma_l,\sigma_{l+1},\Gamma, \Omega_p),$ for $l = 0, \ldots, L-1$, and 
$V_L = V_L(\sigma_L, \Gamma, \Omega_p)$.
The cost $C_{FS}$ is also a function of these parameters.


Given the $C_l, V_l$ for some $\Gamma, \Omega_p$ and a set of shifts, to achieve the optimal cost of Equation (\ref{ml_cost_shifts}), according to the analysis of Giles, the optimal number of samples required at each level is
    \begin{equation}
        N_{l} = \mu \sqrt{\frac{V_{l}}{C_{l}}},~
        \label{num_samples}
    \end{equation}
where the Lagrangian multiplier $\mu$ is given by 
$
\mu = \epsilon^{-2}\left(\sum_{l=0}^{L} \sqrt{V_{l} C_{l}} \right).
$
Then the total variance of the multilevel trace estimation is given by
    \begin{equation}
        V_{FS} = \sum_{l=0}^{L} \frac{V_l}{N_l},
        \label{total_var}
    \end{equation}
and the trace estimator by
    \begin{align}
    \begin{split}
        t(\Gamma \Omega_p D^{-1}) & = \sum_{l=0}^{L-1} \frac{(\sigma_{l+1}-\sigma_{l})}{N_{l}}\sum_{s=0}^{N_{l}} z_{s,l}^{\dag} (D+\sigma_l I)^{-1}\Gamma \Omega_p (D+\sigma_{l+1} I)^{-1}z_{s,l} \\
        & + \frac{1}{N_L} \sum_{s=0}^{N_L} z_{s,L}^{\dag} \Gamma \Omega_p (D+\sigma_L I)^{-1} z_{s,L}
        \label{fs_trace_est}
        \end{split}
    \end{align}
with $z_{s,l}$ and $z_{s,L}$ random vectors with elements from $\{1,-1,i,-i\}$. For the purposes of later discussion regarding the efficacy of our interpolation scheme, we define $V_{total}$ to be the sum of the estimator variance of the levels as
    \begin{equation}
        V_{total} = \sum_{l=0}^{L} V_{l}.
        \label{tosv}
    \end{equation}
Then, our goal is to find the set of shifts $\sigma$ that minimizes Equation (\ref{ml_cost_shifts}) over all sets of shifts. Unfortunately, Equations (\ref{phi_prod}) and (\ref{phi_single}) do not provide a way to determine the shifts analytically.
Moreover, the cost $C_l$ is measured in number of multigrid iterations and, thus, is a discrete function of the shifts.
In the following section, we describe our sampling and interpolation method that is able to predict the variances associated with $V_l$ and $V_L$ and therefore return an approximately optimal set of shifts. Then, we address the question of whether the increase in the number of solves due to multiple levels results in a gain compared to a single level trace estimation at equal target variance.

\section{Methods}
\label{sec:methods}

We describe how the variances corresponding to the functions $V_l$ and $V_L$ of Equations (\ref{phi_prod}) and (\ref{phi_single}) are estimated through sampling, our polynomial interpolation method, how both the number of optimal shifts and their respective values are chosen as a result of the interpolation and our motivation for the choice of probing vectors to use in conjunction with FS.

To facilitate the sampling and interpolation discussion, we define three distinct sets of shifts.
The \textit{sampling} set $\hat{s}$ is a set of $m$ real shifts that are used to solve $m$ linear systems in order to sample the values of $V_l$ and $V_L$ with $\hat{s}_0= 0 < \hat{s}_1 < ... < \hat{s}_{m-1}$. 

The \textit{evaluation} set $s$ is a set of $n$ real shifts, $s_0 = 0 < s_1 < ... < s_{n-1}$, where we evaluate the interpolating polynomials to obtain a prediction of the variances $V_l$ and $V_L$ at each pair of values $s_i$, $s_j$ or each value $s_i$ respectively. The discretization is chosen to capture the shape of the manifold to be interpolated. $s_{n-1}$ is chosen to be a value well within the interior of the spectrum of $D$. The evaluation set contains the sampling set.

The \textit{optimal} set is the set of $L$ shifts chosen from $s$ that minimize Equation (\ref{ml_cost_shifts}) over all choices of $s_i$, i.e., $\sigma_0 = s_0 = 0 < \sigma_1 = s_{j_1} < ... < \sigma_L = s_{j_L}$ with
    \begin{equation}
    \sigma = \argmin_{1 \leq j_1<j_2<...<j_L\leq n} C_{FS}(s_0,s_{j_1},\ldots ,s_{j_L}).
    \label{opt_shifts}
    \end{equation}

\subsection{Sampling the Variance}
\label{subsec:sample}

Given our set of sampling shifts $\hat{s}$, we wish to estimate the variances \begin{align}
    \label{sampled_prod}
   V_l(\hat{s}_i,\hat{s}_j,\Gamma,\Omega_p) & =
   (\hat{s}_j-\hat{s}_i)^2 \bar V_l(\hat{s}_i,\hat{s}_j,\Gamma,\Omega_p)\\
    V_L(\hat{s}_i, \Gamma, \Omega_p) & = \Var[ t( \Gamma \Omega_p (D+\hat{s}_i I)^{-1} ) ],
    \label{sampled_single}
\end{align}
for $i = 0,\ldots,m-1$ and $j = i,\ldots,m-1$. Here, we introduce $\bar V_l$ as
\begin{equation}
    \bar V_l(\hat{s}_i,\hat{s}_j,\Gamma,\Omega_p) = \Var[t((D+\hat{s}_i I)^{-1} \Gamma \Omega_p (D+\hat{s}_j I)^{-1} )],
    \label{bar_vl}
\end{equation}
which will be sampling instead of $V_l$ since $V_l$ decreases quadratically with the term $(\hat{s}_j-\hat{s}_i)$ and therefore poses difficulties for interpolation that the sampling of $\bar V_l$ alleviates. These difficulties are discussed in Section \ref{subsec:interp}.

The variance of the trace estimator $t(A)$ cannot be computed in practice using Equation (\ref{test_var}) but is instead estimated via
\begin{equation}
    Var[t(A)] = \frac{1}{N_s}\sum_{k=1}^{N_s}(z_k^{\dag}Az_k)^*(z_k^{\dag}Az_k)-t(A)^*t(A).
    \label{practical_variance}
\end{equation}
If the vectors $z_k$ are random vectors with elements $\{1,-1,i,-i\}$, then Equation (\ref{practical_variance}) is an estimator of Equation (\ref{test_var}), i.e., of the Frobenius norm of the off-diagonal elements of $A$. Thus, to estimate the variance in Equation (\ref{bar_vl}), we compute Equation (\ref{practical_variance}) with $A = (D+\hat{s}_i I)^{-1} \Gamma \Omega_p (D+\hat{s}_j I)^{-1}$ for $i = 0,...,m-1$ and $j = i,...,m-1$. In the case of Equation (\ref{sampled_single}), we compute Equation (\ref{practical_variance}) with $A = \Gamma \Omega_p (D+\hat{s}_i I)^{-1}$ for $i = 0,...,m-1$.

Computationally, Equations (\ref{bar_vl}) and (\ref{sampled_single}), require the solution of systems of linear equations of the form
\begin{align}
      \label{shifted_linear_systems}
      (D+\hat{s}_i I)x &= z \\
      (D+\hat{s}_i I)^{\dag}y &= z.
      \label{conj_shifted_linear_systems}  
\end{align}
While this procedure can be done for a general random noise vector $z$, we employ the use of full spin-color dilution \cite{wilcox_partition,FOLEY2005145} and probing, therefore $z$ has support on lattice sites specified by the coloring of the lattice and for a single spin-color index. For details on the parameters of the linear solves, such as configuration details, see Sec. \ref{sec:results}. Additionally, due to the $\gamma_5$-hermiticity of $D$, the use of full spin-color dilution allows us to obtain the solutions of Equation (\ref{conj_shifted_linear_systems}) from the solutions of Equation (\ref{shifted_linear_systems}). This results in $12cmN_s$ total inversions, where $c$ is the number of probing vectors and the factor of 12 arises from the use of spin-color dilution. The algorithm for computing the variances $V_L$ and $\bar V_l$ is given below for general noise vectors, $z$, as the extension to using spin-dilution and probing vectors is straight forward. \\
\newline
\begin{small}
\noindent \textbf{Input}: The $m$ sampling shifts, $\hat{s}$, operators $D$, $\Gamma$, $\Omega_p$, and $N_s$ noise vectors, $z$.
\newline
\textbf{Output}: Sampled variances $V_L(\hat{s}_i, \Gamma, \Omega_p)$ and $\bar V_l(\hat{s}_i, \hat{s}_j, \Gamma, \Omega_p)$.
\begin{enumerate}  
\item For $k = 0:N_s-1$
\item \hspace{0.3cm} For $i = 0:m-1$
\item \hspace{0.6cm} Solve $(D+\hat{s}_i I)x_k^{(i)} = z_{k}$
\item \hspace{0.6cm} Compute $y_k^{(i)} = \gamma_5 x_k^{(i)}$ as the solution of Eqn. (\ref{conj_shifted_linear_systems})
\item \hspace{0.3cm} For $i= 0:m-1$
\item \hspace{0.6cm}
Compute $z_k^{\dag}Az_k = z_k^{\dag}\Gamma \Omega_p x_k^{(i)}$ with $A = \Gamma \Omega_p (D+\hat{s}_i I)^{-1}$
\item \hspace{0.6cm}
Update $V_L(\hat{s}_i)$ estimation from Eqn. (\ref{practical_variance})

\item\hspace{0.6cm} For $j = i:m-1$
\item \hspace{0.9cm} 
Compute $z_k^{\dag}Az_k = y_k^{(i){\dag}} \Gamma \Omega_p x_k^{(j)}$ with $A = (D+\hat{s}_i I)^{-1} \Gamma \Omega_p (D+\hat{s}_j I)^{-1}$

\item \hspace{0.9cm} Update $\bar V_l(\hat{s}_i,\hat{s}_j)$ estimation from Eqn. (\ref{practical_variance}) \\
\end{enumerate}
\end{small}
Given this large number of inversions, a judicious choice of the number of samples, $N_s$, is required that is both practical to compute and gives an accurate estimate of the true variance. In \cite{doi:10.1137/17M1137541}, it was shown that accurate order of magnitude estimate of the Frobenius norm of a matrix can be obtained with a small number of samples. Since the calculation of the variance of the trace estimator is in fact an estimation of the Frobenius norm of the off-diagonal elements, we can obtain an accurate estimate of the variance of the trace estimators given by Equations (\ref{bar_vl}) and (\ref{sampled_single}) with just a few samples. We therefore sample the variance with $N_s = 5$ samples in order to keep the expense of sampling practical. The number of inner products required to compute $V_L$ and $\bar V_l$ scales as  $O(12 c m^2 N_s)$, so their cost is negligible compared with the cost of $O(12 c m N_s)$ inversions in the sampling phase. 
 
Since the total cost of sampling  scales linearly with $m$, the number of shifts in the sampling set $\hat{s}$, a fine enough discretization of $\hat{s}$ from which to obtain the optimal shifts $\sigma$ through sampling alone is too expensive. This is the reason why we pick a small set of sampling shifts and produce the larger evaluation set $s$ through interpolation.
The choice of the sampling set of shifts is motivated by the following factors: In order for Equation (\ref{tele_sum_prod_disp}) to hold, $\hat{s}_0 = 0$ is required. We also choose the largest shift $\hat{s}_m$ to be a value well within the interior of the spectrum of $D$, as the variance of the trace estimator of $(D+\hat{s}_{m-1}I)^{-1}$ will be small due to the decay coefficient of the inverse, as inferred from Figure \ref{rho_v_lambda}. Additionally, the behavior of $\bar V_l(\hat{s}_i,\hat{s}_j)$ as a function of $\hat{s}_i$ motivates our choice for the first non zero sampled shift, $\hat{s}_1$. In Figure \ref{boundary}, we observe that the sampled points of $\bar V_l$, given by red bursts, decrease rapidly for small shifts away from 0. To accurately capture this rapidly changing area with interpolation, we choose $\hat{s}_1$ to be a value near zero. The other values of $\hat{s}$ can be chosen equidistant in log space, as it is observed that $\bar V_l$ exhibits near exponential decay for shifts larger than 0.15. This motivates our choice to choose $m = 5$ sampling shifts in order to keep the expense of sampling practical, with $\hat{s} = \{0.0, 0.05, 0.25, 0.50, 1.00\}$.

\comment{Once Equation (\ref{shifted_linear_systems}) has been solved for our set of sampling shifts using $N_s$ samples, it is only a matter of performing the inner products in order to estimate the variances associated with $V_l$ and $V_L$, which are given by
\begin{align}
    \label{sampled_prod}
    V_l(\Gamma,\Pi_p,\hat{s}_i,\hat{s}_j) & = \Var[t((D+\hat{s}_i I)^{-1} \Gamma \Pi_p (D+\hat{s}_j I)^{-1} )],\\
    V_L(\Gamma,\Pi_p,\hat{s}_i) & = \Var[ t( \Gamma \Pi_p (D+\hat{s}_i I)^{-1} ) ].
    \label{sampled_single}
\end{align}
It is prudent to mention that Equation (\ref{sampled_prod}) does not include the prefactor of the shift difference, in contrast to Equation (\ref{phi_prod}). Our justifications for not including it are explained in Sec. \ref{subsec:interp}.
The inner products are performed for $i = 0,...,m-1$ and $j \geq i$ for Equation (\ref{sampled_prod}) and $i = 0,...,m-1$ for Equation (\ref{sampled_single}) in order to obtain the variance of the estimators. For our choice of $m = 5$, this results in 15 sampled values for $V_l$ and 5 sampled values for $V_L$.} 

We also remark that the variance of Equation (\ref{sampled_single}) can be reduced when the value of the bare quark mass parameter, $m_q$, is large through the use of the Generalized Hopping Parameter Expansion \cite{G_lpers_2014}\cite{FreqSplit}. We employ the use of the Generalized Hopping Parameter Expansion for values of $\hat{s}_i$ when $m_q+\hat{s}_i > 0$, truncating the expansion at an order of $k = 4$. Even though it is not required in the sampling stage, we remark that the trace of the terms of order $k < 4$ can be calculated exactly with the use of distance-3 displacement-$p$ probing vectors, where $p$ corresponds to the size of the displacement from $\Omega_p$. Having completed this sampling procedure, we have all the necessary values to interpolate the sampled values of $\bar V_l$ and $V_L$ to obtain predicted variances using our evaluation set of shifts, $s$.

\subsection{Interpolation}
\label{subsec:interp}
\begin{figure}[!h]
\centering
\begin{minipage}{0.6\textwidth}
\noindent \textbf{Input}: The sampled variances $\bar V_l$, the $m$ sampled shifts, $\hat{s}$ and $n$ evaluation shifts, $s$.
\newline
\textbf{Output}: Predicted variances $V_l$.
\begin{enumerate}
\item for $j = 1:m-1$
\item \hspace{0.5cm}$w = ln(\bar V_l(\hat{s}_i,\hat{s}_j))$~~~~with $i = 0,\ldots,j$
\item \hspace{0.5cm}$q = Interpolate(\hat{s}_{0,\ldots,j},w)$
\item \hspace{0.5cm}for $k = 0:n-1$
\item \hspace{1.0cm}$\bar V_l(s_k,\hat{s}_j) = e^{q(s_k)}$
\item \hspace{0.5cm}end
\item end
\item $w = ln(\bar V_l(\hat{s}_i,\hat{s}_j))$~~~~with $i=j=0,\ldots,m-1$
\item $q = Interpolate(\hat{s}_{0,\ldots,m-1},w)$
\item for $k = 0:n-1$
\item \hspace{0.5cm}$\bar V_l(s_k,s_k) = e^{q(s_k)}$
\item end
\item for $i=0:n-1$
\item \hspace{0.5cm}$w = ln(\bar V_l(s_i,\hat{s}_j))$~~~~with $j = 0,\ldots,m-1$
\item \hspace{0.5cm}$q = Interpolate(\hat{s}_{0,\ldots,m-1},w)$
\item \hspace{0.5cm}for $k = 0:n-1$
\item \hspace{1.0cm}$ V_l(s_i,s_k) = (s_k-s_i)^2 e^{q(s_k)}$
\item \hspace{0.5cm}end
\item end
\end{enumerate}
\label{alg:interp}
\end{minipage}
\hfill
\begin{minipage}{0.35\textwidth}
\includegraphics[width=0.75\textwidth,scale=0.5]{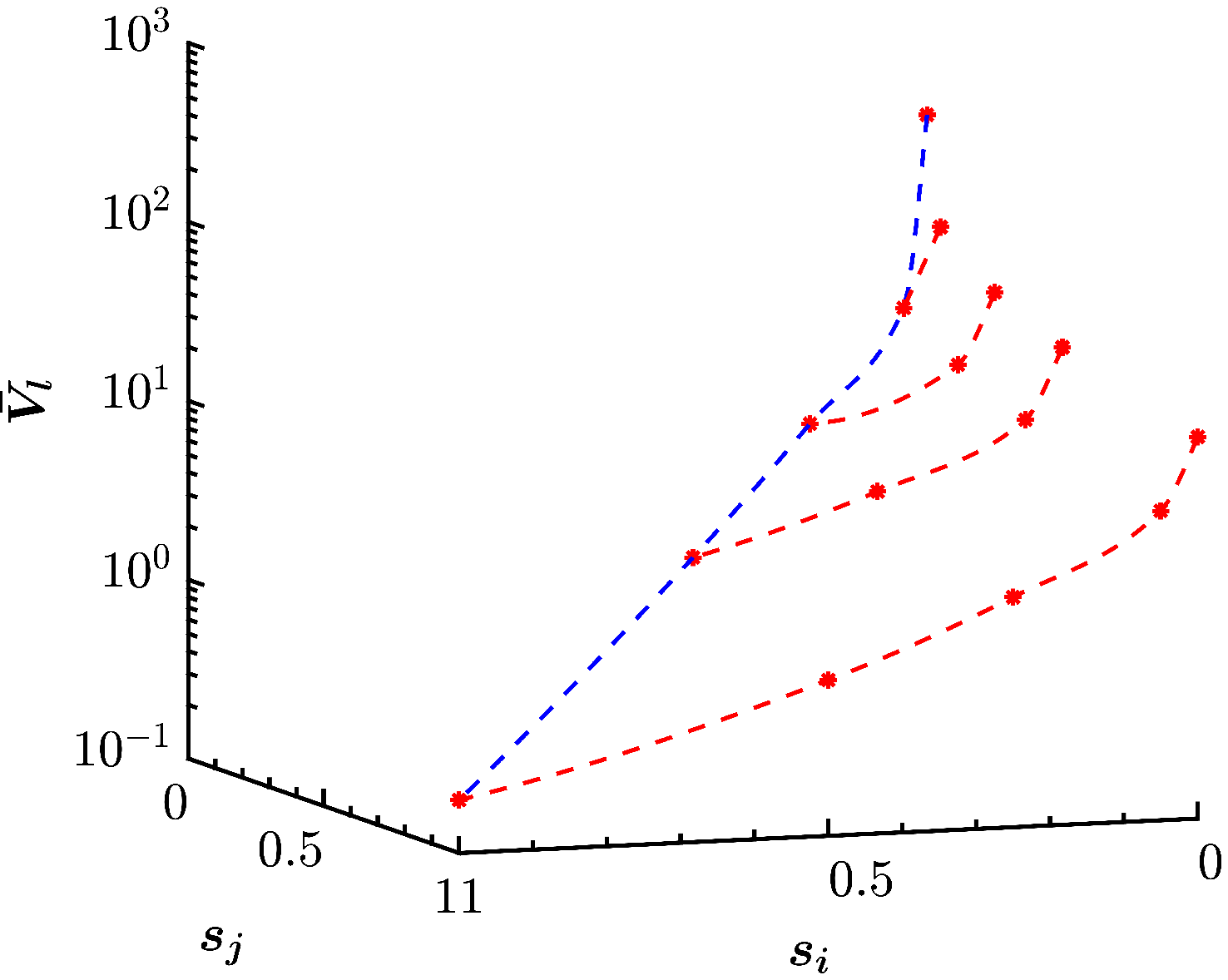}
\includegraphics[width=0.75\textwidth,scale=0.5]{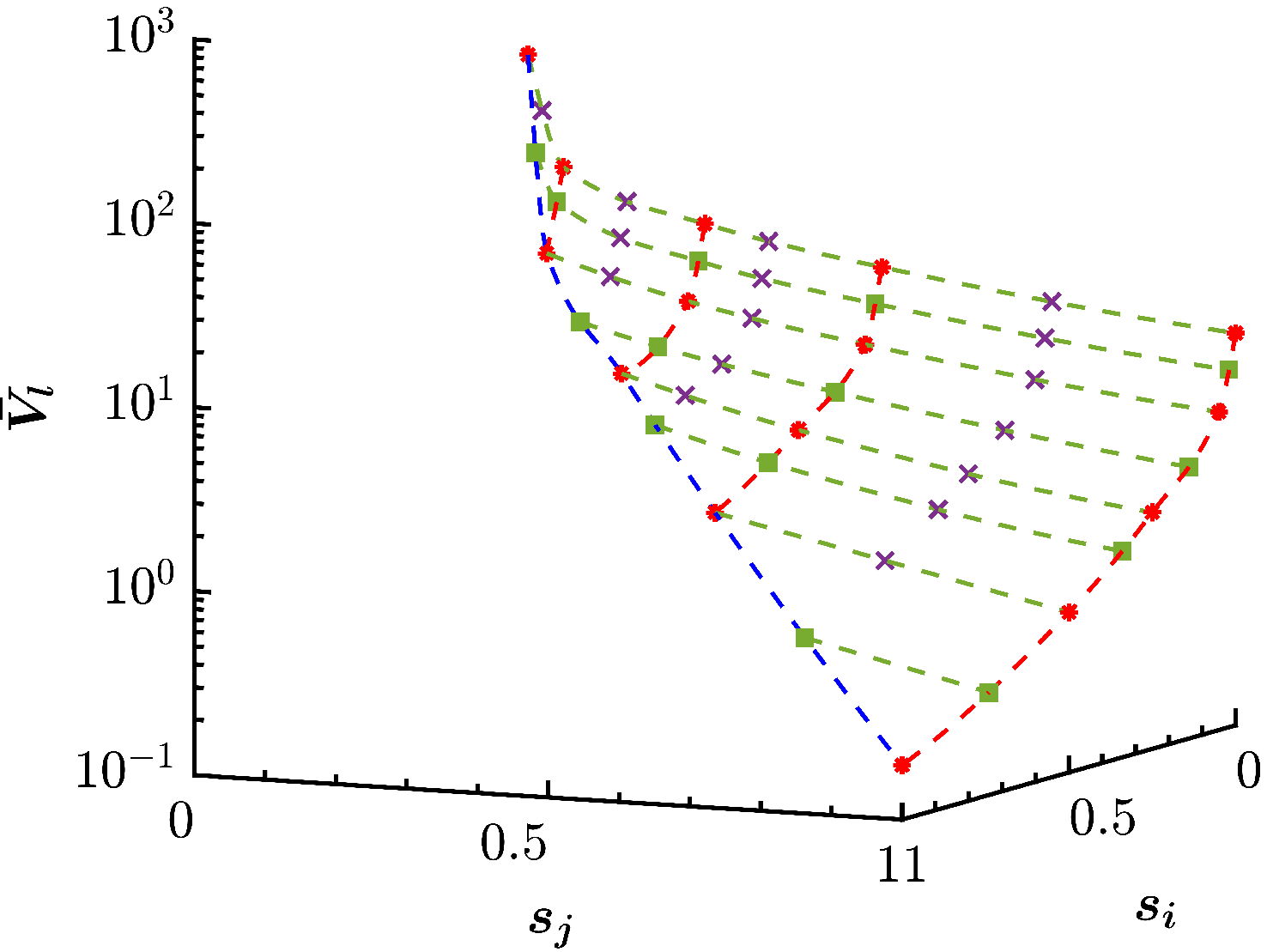}
\includegraphics[width=0.75\textwidth,scale=0.5]{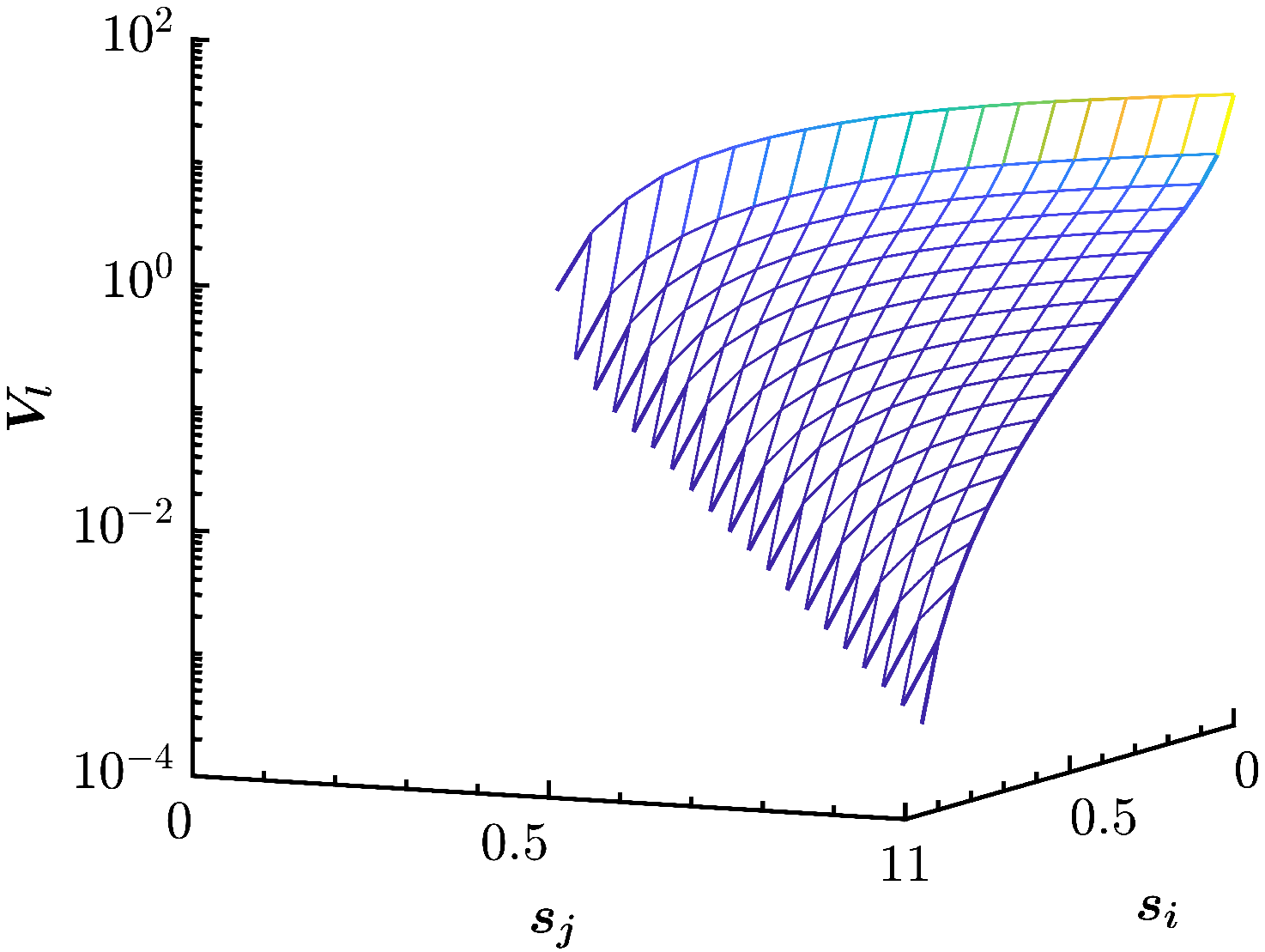}
\end{minipage}
\caption{(Left) The algorithm for the $2D$ interpolation of $\bar V_l$. (Top right) The sampled variances $\bar V_l(\hat{s}_i,\hat{s}_j)$ (red bursts) for $\hat{s}_i, \hat{s}_j \in \{0.0, 0.05, 0.25, 0.5, 1.0\}$. The red dashed lines denote the polynomials generated at fixed $\hat{s}_j$ at step $3$ of the algorithm. The blue dashed line denotes the polynomial generated when $\hat{s}_i = \hat{s}_j$ at step $9$ of the algorithm. (Right Center) Rotated 45$^\circ$ with respect to top subgraph and shown for an evaluation set of size $n = 9$. Green squares denote the evaluated points of polynomials generated at fixed $\hat{s}_j$ and $\hat{s}_i = \hat{s}_j$ at steps $(4-6)$ and $(10-12)$. These evaluated points can then be used to form interpolating polynomials at fixed $s_i$, denoted by green dashes, at step $15$. Purple crosses denote the evaluated points of the polynomials generated at fixed ${s}_i$ in step $17$. (Right bottom) The final manifold of $V_l$ for $n = 20$.}
\label{boundary}
\end{figure}

Once the sampled values of $\bar V_l$ and $V_L$ have been calculated, we are now in a position to interpolate the values to obtain a much larger ``shift space" to explore in order to find the set of shifts $\sigma$ from our evaluation set $s$ that approximately minimizes Equation (\ref{ml_cost_shifts}). While the values of $V_L$ are simple to interpolate, the values of $\bar V_l$ are more difficult due to the two dimensional dependence on the sampling shifts. A naive two dimensional interpolation and interpolating the variances of $V_l$ directly instead of $\bar V_l$ resulted in poor accuracy. This is especially true for small shift differences where the $V_l$ vanishes rapidly, as shown in the bottom right graph of Figure \ref{boundary}. Moreover, without the shift difference term, $\bar V_l$ displays near exponential decay with larger shifts, as shown in the top two subgraphs of Figure \ref{boundary}, which can be captured easily by interpolating in log space. The above two reasons motivated us to sample values of $\bar V_l$ as defined in Equation (\ref{bar_vl}).

Because of the monotonic decrease of $\bar V_l$ with larger shifts, we use piecewise cubic Hermite interpolating polynomials (PCHIP) \cite{pchippub} from the Boost library \cite{boost} in order to interpolate the variances corresponding to $\bar V_l$ and $V_L$, as PCHIP preserves monotonicity and continuity (see  \cite{pchippub} for details). First, we use the sampled points $ln(V_L(\hat{s}_i))$ to form the piecewise cubic Hermite interpolating polynomial, $q$, on the interval $[0, \hat{s}_{m-1}]$.
 
Then, we can predict the variance $V_L(s_k)$ for all $n$ evaluation shifts $s_k$ as
\begin{equation}
    V_L(s_k) = e^{q(s_k)}.
\end{equation}

The interpolation of $\bar V_l$ is also performed in logarithmic scale. However, due to the two dimensional dependence of $\bar V_l$ on the sampling set, the interpolation of $\bar V_l$ is performed first along the first dimension, producing $V_l(s_k,\hat{s}_j)$ for all $s_k$, then along the diagonal of the two dimensions producing all $V_l(s_k,s_k)$, and finally along the second dimension using all sampled or produced points up to now to interpolate any arbitrary $V_l(s_k,s_j)$. Figure \ref{boundary} gives the algorithm for the two dimensional interpolation of $\bar V_l$ and the right subgraphs illustrate the process. We remark that when only two sampled points of $\bar V_l$ are available (e.g. in step 3 of the algorithm with $j = 1$) PCHIP cannot be used. In such cases, linear interpolation is used instead. 







Lastly, we address how the interpolation of the level costs is performed. Specifically, these are the predicted costs of the solution of the systems (\ref{shifted_linear_systems}) and (\ref{conj_shifted_linear_systems}) for each evaluation shift $s_i$. 
The cost of the inner products does not change with the choice of shift and therefore is not considered in the optimization.
The solver costs at each sampled point $\hat{s}_i$ are given by 
    \begin{equation}
        C(\hat{s}_i) = 12 c r(\hat{s}_i)
        \label{solver_cost_shift}
    \end{equation}
where $c$ is the number of colors used in probing and $r(\hat{s}_i)$ is the number of iterations required for the convergence of linear equations at the shift $\hat{s}_i$. A hallmark of multigrid solvers in LQCD is that the number of fine grid iterations remains small or increases only slightly as the quark mass approaches zero \cite{PhysRevLett.105.201602,PhysRevD.97.114513,Whyte2020DeflatedGW}. 
Thus, as the lattice Dirac operator is shifted away from zero, 
$r({s}_i)$ decreases slightly in a piecewise fashion.  We use next neighbor interpolation to obtain the solver cost at each evaluation point,
$r(s_i) = r(\hat{s}_j)$, where 
$\hat{s}_{j-1} < s_i \leq \hat{s}+j$.
We remark that next nearest neighbor communication guarantees not to overestimate the cost of the solver, and thus not to underestimate the number of $N_l$ iterations at each level.

\subsection{Shift Selection}
\label{subsec:shift_select}
\begin{figure}[h!]
\begin{center}
\mbox{\subfigure{\includegraphics[width=0.45\textwidth,scale=1.0]{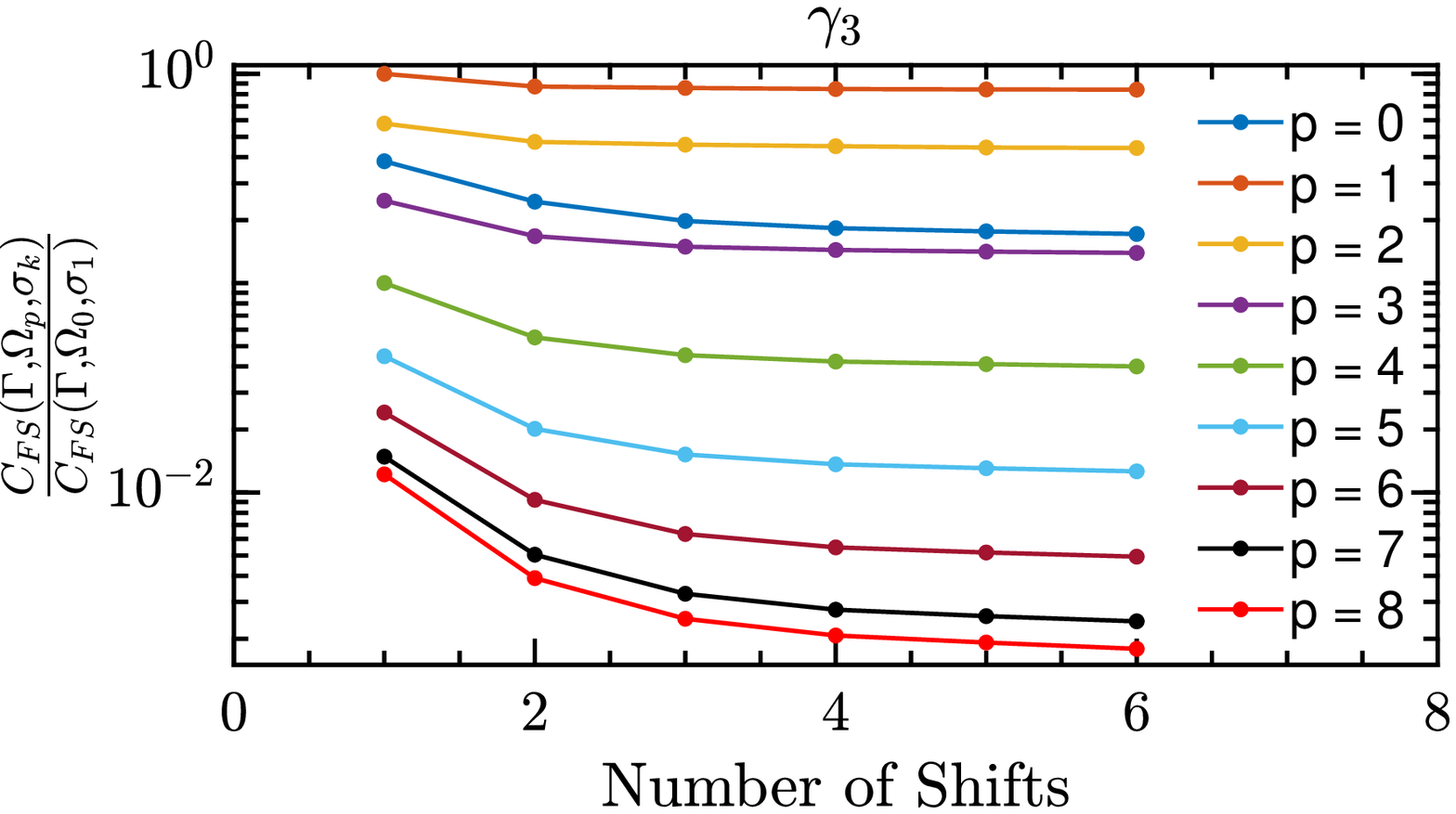}}

\hspace{0.0cm}

\subfigure{\includegraphics[width=0.45\textwidth,scale=1.0]{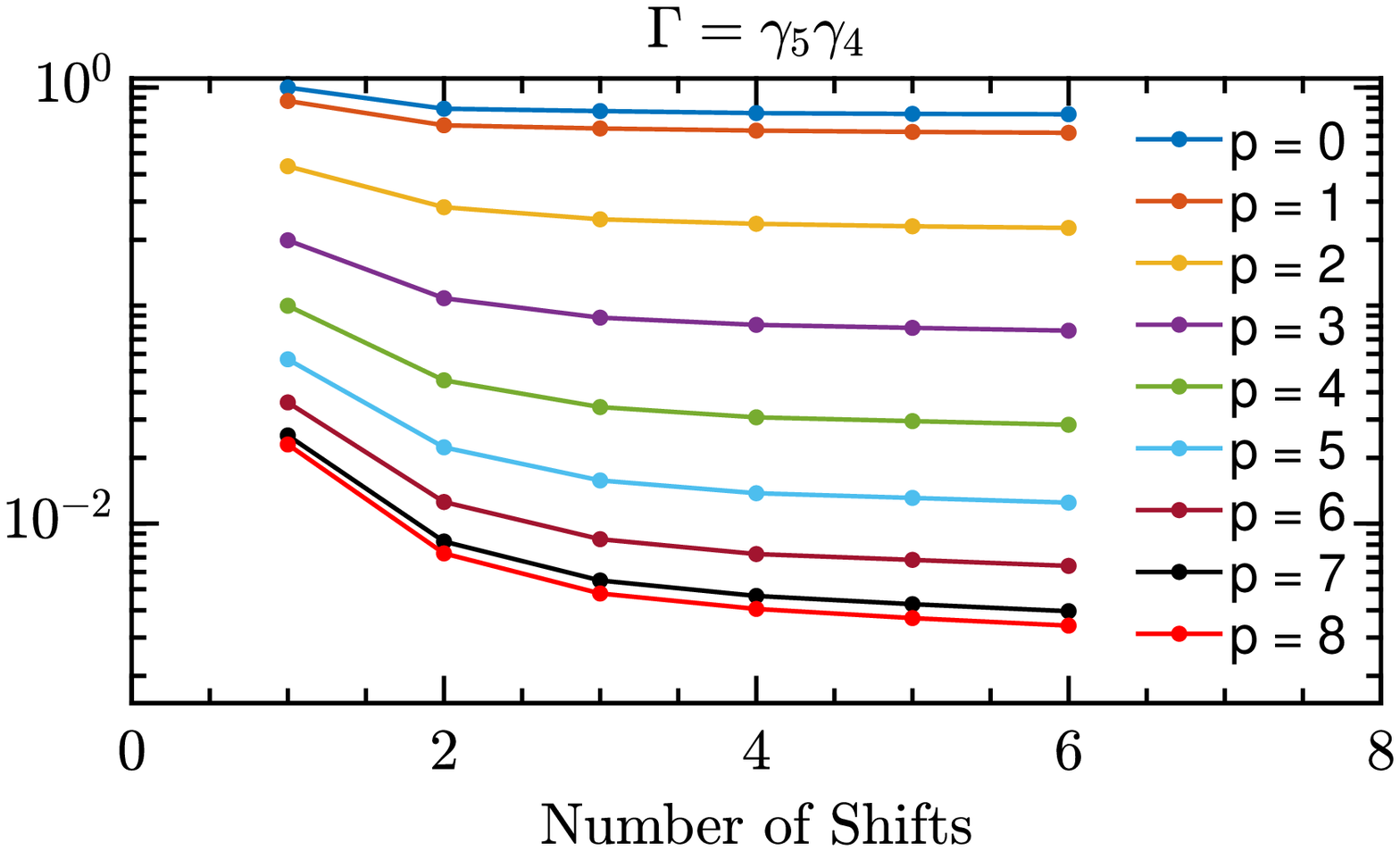} }}

\caption{The normalized predicted minimum cost for $\gamma_3$ and $\gamma_5 \gamma_4$ for all displacements as a function of the number of shifts, where $p$ is the size of the z-displacement.}
\label{pred_mincost_v_num_shifts}
\end{center}
\end{figure}

The combination of all shifts used in the interpolation gives rise to an $L$-dimensional manifold of the multilevel cost in Equation (\ref{ml_cost_shifts}), where $L$ is the number of levels. To find the values of $\sigma$ corresponding to the minima of the manifold given by Equation (\ref{opt_shifts}) we iterate over all allowed combinations of the $n$ shifts used for evaluation of the interpolating polynomials. The value of $n$ has to then remain somewhat modest. It is also important to properly choose how many shifts (or levels) to use in the multilevel trace estimation. We therefore examine the predicted minimum cost as a function of the number of shifts at equal target variance $\epsilon^2$. Figure \ref{pred_mincost_v_num_shifts} displays the normalized predicted multilevel cost, $\frac{C_{FS}(\Gamma, \Omega_p, \sigma_k)}{C_{FS}(\Gamma, \Omega_0,  \sigma_1)}$. It can be observed that the cost rapidly begins to plateau and very little gain can be made beyond five or six shifts. Thus, we choose to use six shifts (e.g. seven levels) for the duration of this work. 

Initial attempts at interpolation focused heavily on shifts near zero due to the large variance associated with $\Gamma \Omega_p D^{-1}$. This is because by splitting the inverse more finely at lower shifts, large variances near zero are suppressed by the effect of $(\sigma_j-\sigma_i)^2$. However, experimentation with the values of $s$ show that, although this is true in general, having a very fine discretization extremely close to zero is unnecessary.

\begin{figure}[!h]
\begin{center}
\mbox{\subfigure{\includegraphics[width=0.30\textwidth,scale=1.0]{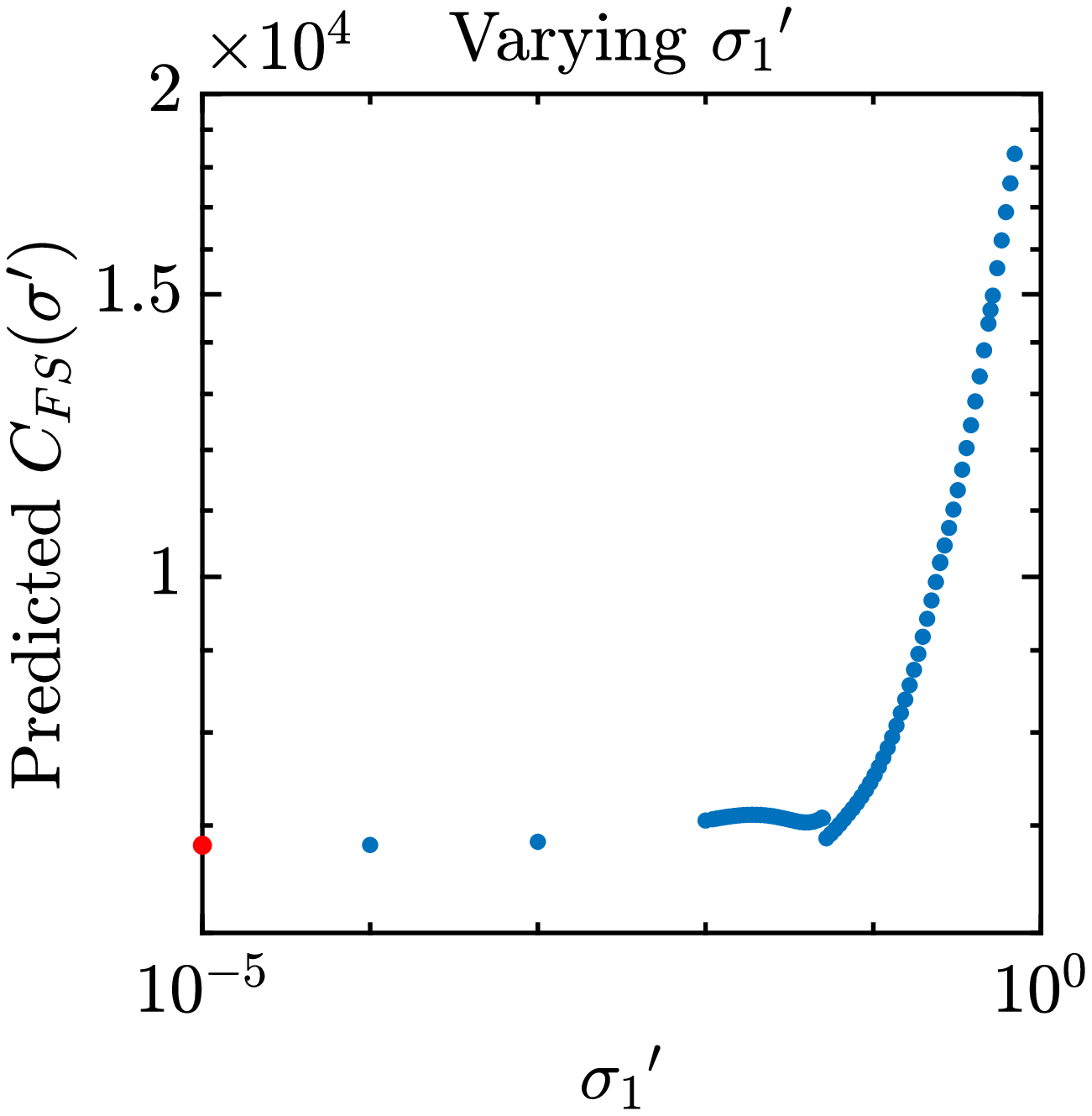}}

\subfigure{\includegraphics[width=0.30\textwidth,scale=1.0]{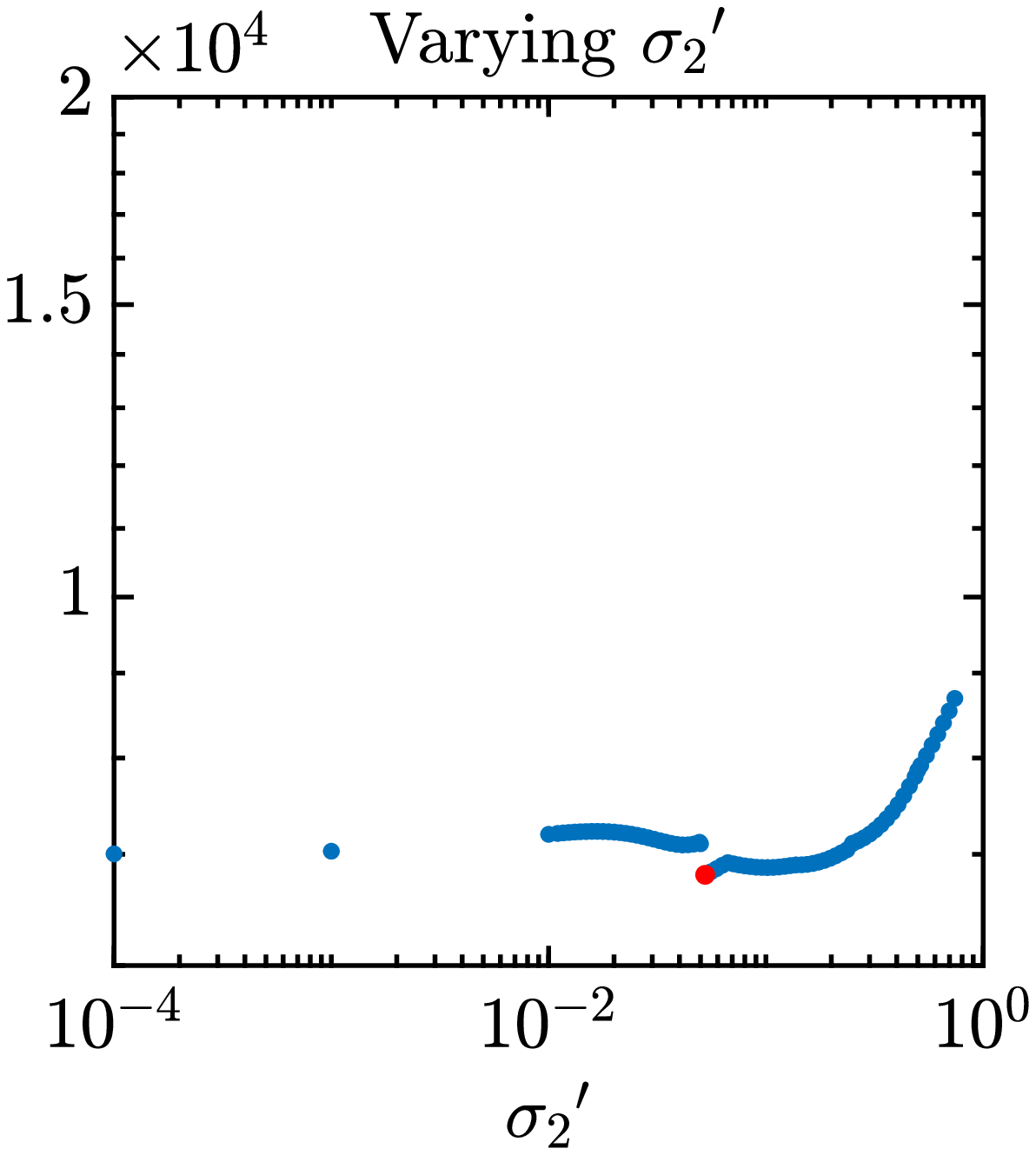} }

\subfigure{\includegraphics[width=0.30\textwidth,scale=1.0]{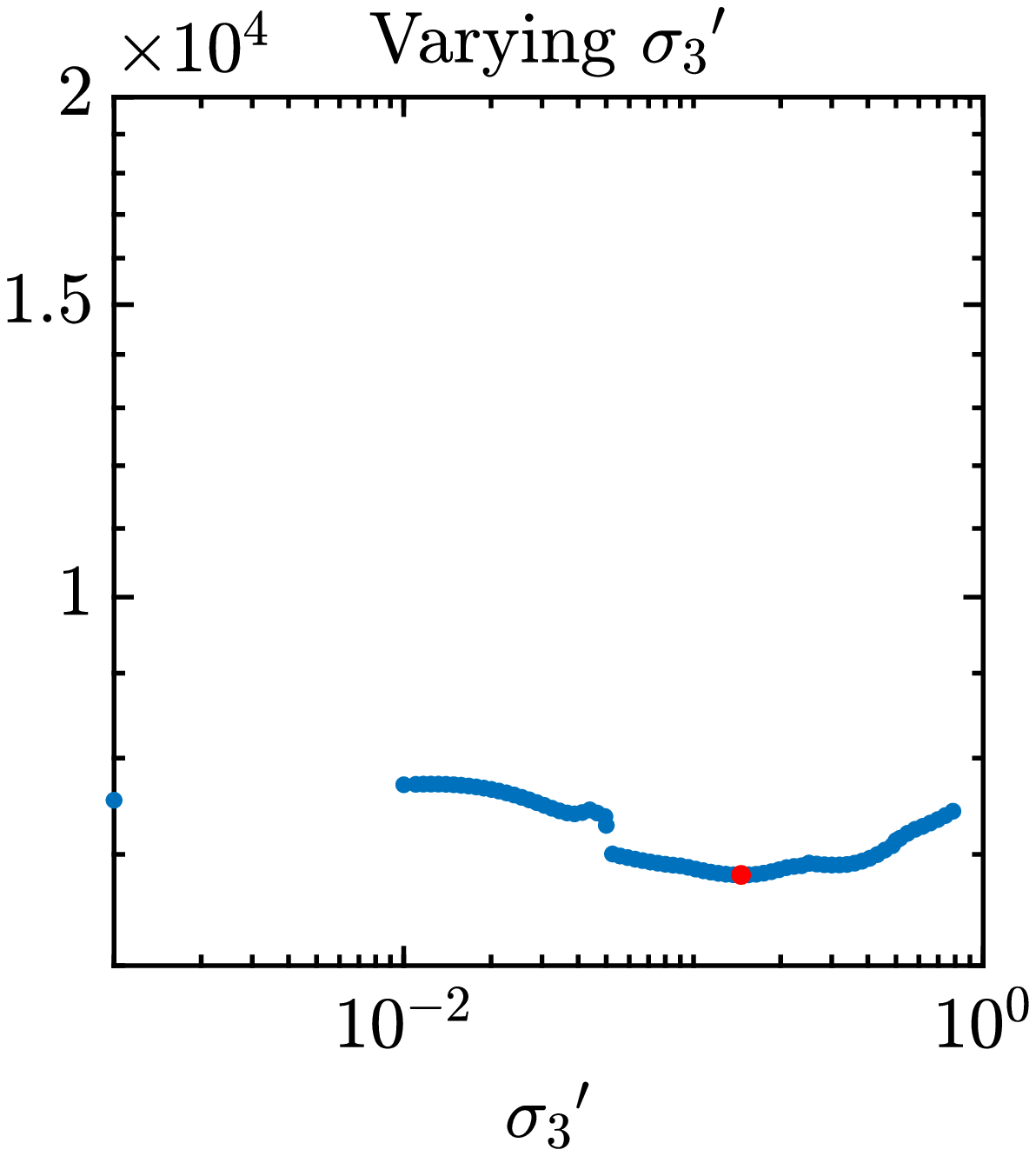}}}

\mbox{\subfigure{\includegraphics[width=0.30\textwidth,scale=1.0]{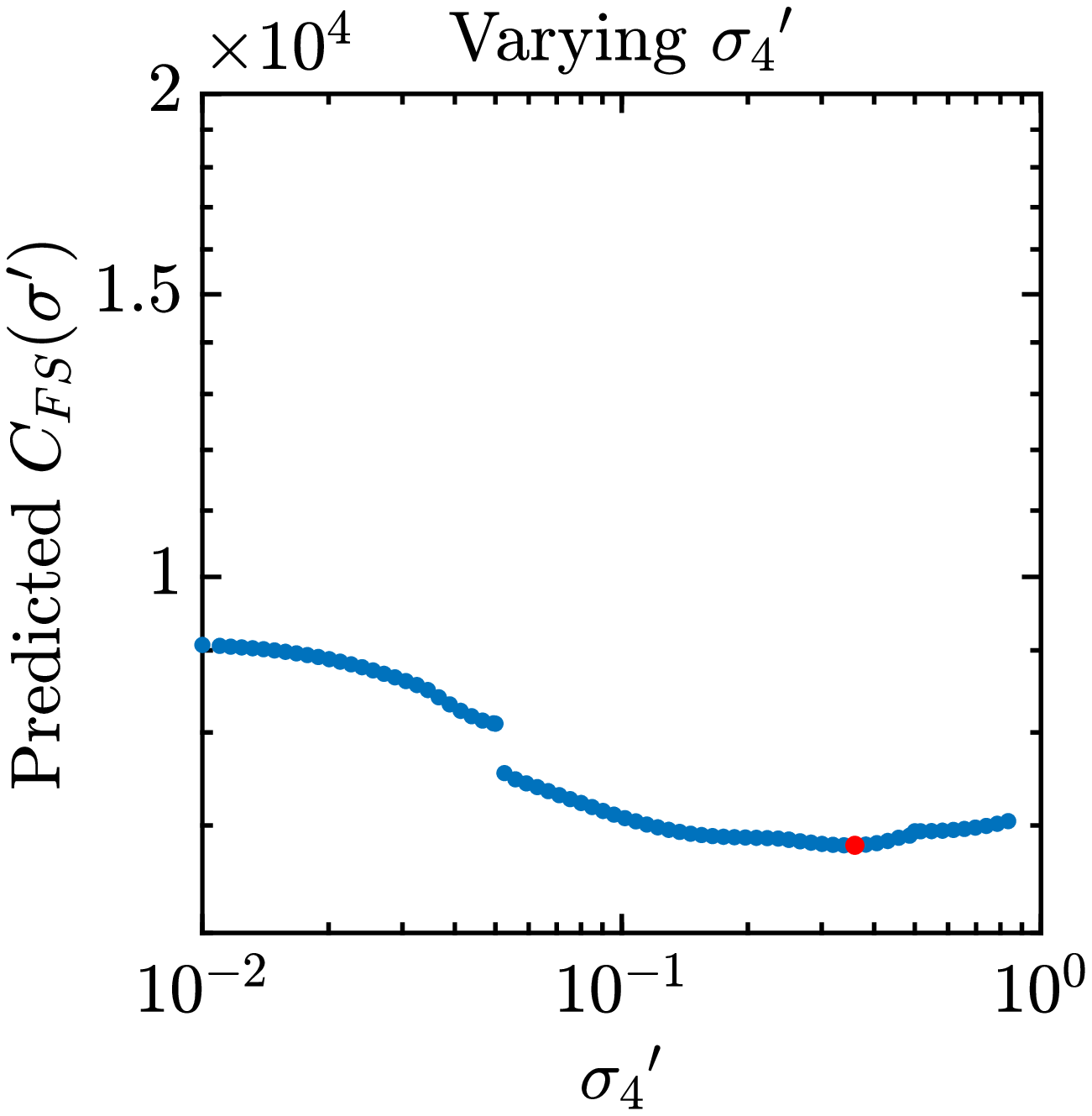}}

\subfigure{\includegraphics[width=0.30\textwidth,scale=1.0]{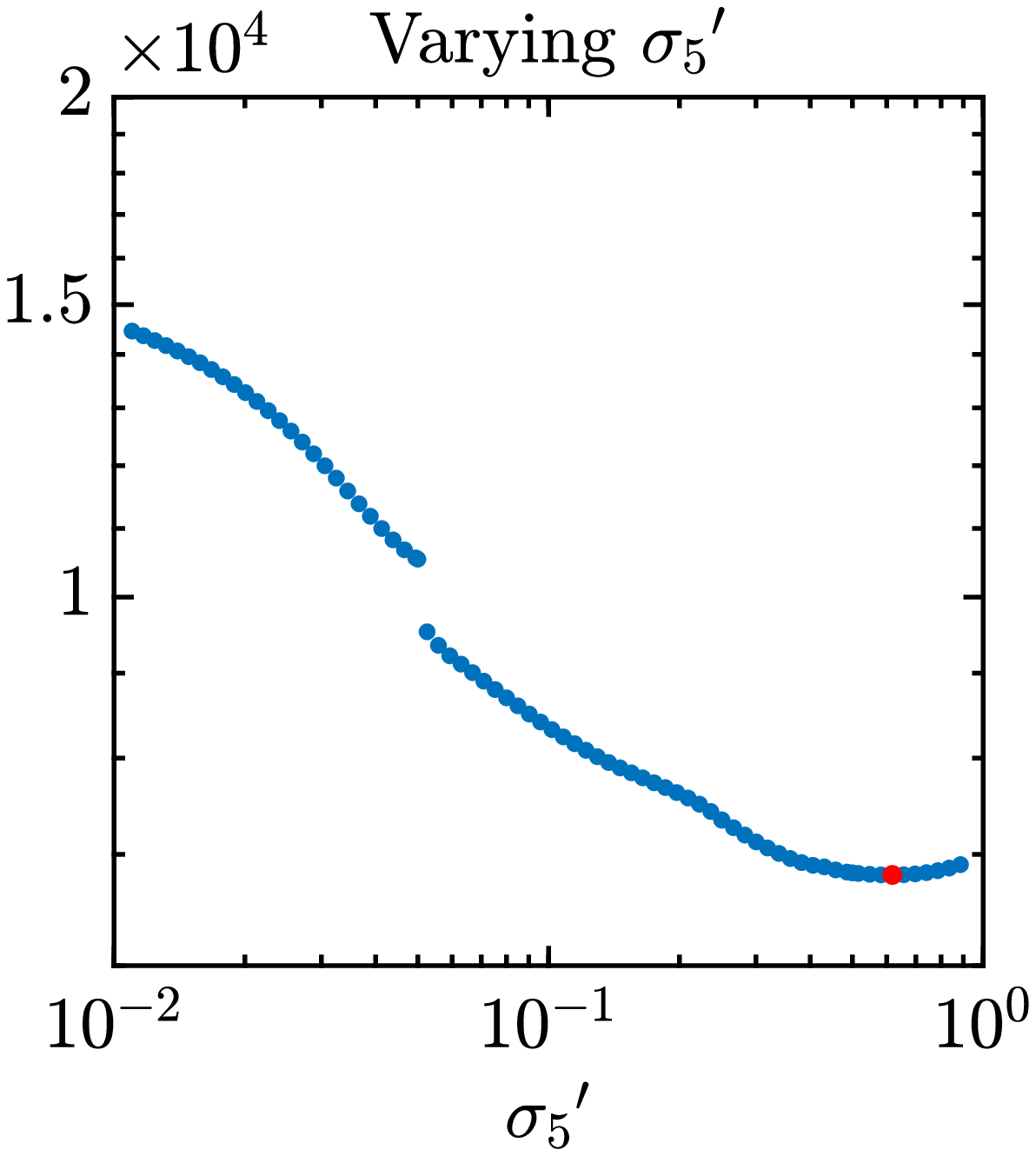} }

\subfigure{\includegraphics[width=0.30\textwidth,scale=1.0]{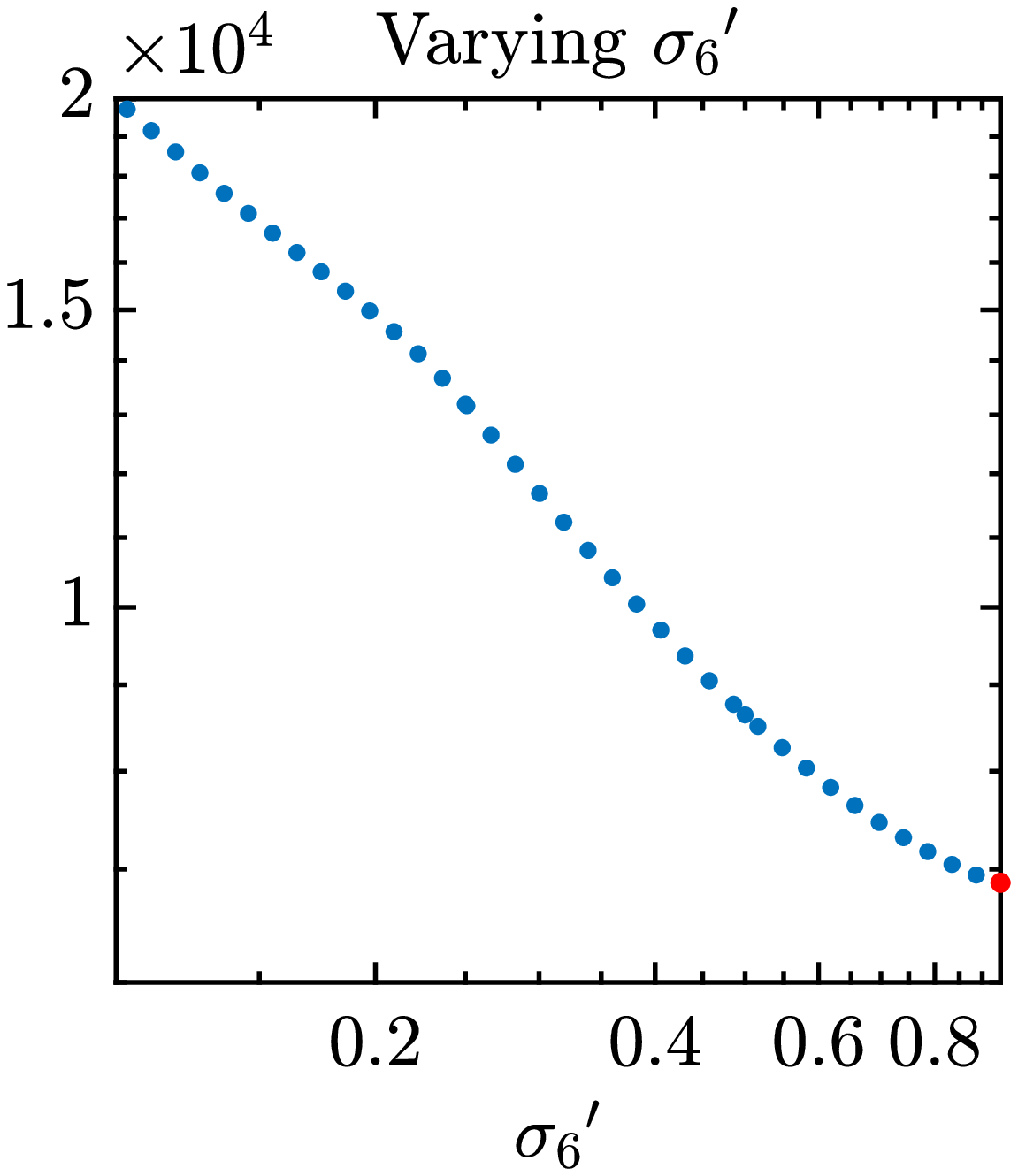}}}

\caption{The predicted multilevel cost $C_{FS}$ as a function of each $\sigma_i'$. Each $\sigma_i'$ is fixed to a value of $s$ and the other five take on the values that minimize the predicted multilevel cost for that value of $s$, to create a new set of shifts, $\sigma'$. The values where $\sigma_i' = \sigma_i$, given by Equation (\ref{opt_shifts}), are given in red.}
\label{si_min}
\end{center}
\end{figure}

To examine this behavior, we explore the region around the minimum multilevel cost given by the set of shifts $\sigma$ from Equation (\ref{opt_shifts}). We vary each $\sigma_i$ where $i = 1,\ldots,6$ to take on values of $s_k$ for all $k$ such that $1 \leq j_{i-1} < k < j_{i+1} \leq n-1$. The other five elements of $\sigma$ take on values to create a new set of six shifts $\sigma{'}$ that minimizes the multilevel cost such that $0 \leq \sigma_{i-1}' <\sigma_i'= s_k< \sigma_{i+1}' \leq s_{n-1}$. This allows us to examine the predicted multilevel cost as a function of each $\sigma_i'$ as shown in Figure \ref{si_min}. Effectively, each plot shows a slice of the six dimensional manifold of the multilevel cost. We observe that it is much more important to finely discretize larger shifts than it is to finely discretize smaller shifts. Examining the cost as a function of $\sigma_1'$ (upper left) shows that the cost displays a very gradual increase for shifts near zero, so we expect the contribution of $C_0V_0$ to the multilevel cost given by Equation (\ref{ml_cost_shifts}) to be very small for small values of $s$. It is therefore much less important to finely discretize the region corresponding to small values of $s$. The other five shifts display much more pronounced minima and it is therefore more important to finely discretize that region. Thus, to evaluate the interpolating polynomials we use a shift discretization $s = $ {\tt [logspace(-5,-2,4)~logspace(log10(1e-2+1e-3),0,76)]} (MATLAB notation), which focuses on shifts larger than 0.1. 
Finally, we mention that the discontinuities in the multilevel cost seen in Figure \ref{si_min} are due to the piecewise, discontinuous interpolation of the solver costs and the small number of fine grid iterations that multigrid takes. If a solver with more iterations were chosen, the cost interpolation would have more granularity and this discontinuity would not be present.

\subsection{Selection of Probing Vectors}
The choice of probing vectors is independent from the above issues and is motivated by the following factors: The number of colors should remain relatively small to control the cost in Equation (\ref{solver_cost_shift}), as the same set of probing vectors is used for all levels. Due to the approximate exponential decay of $D^{-1}$, the trace rapidly decays as a function of the displacement in the lattice. This leads to large relative error in the estimation of the trace for large displacements \cite{switzer2021probing}. The probing vectors selected should then reduce the variance at large displacements of the lattice such that the relative error of the trace remains small. 

Based on the above, we examine three sets of (displacement-$p$, distance-$k$) probing vectors: $p4k4$, $p6k7$ and $p8k7$, having 14, 37 and 16 colors, respectively, and generated with \cite{heather_coloring}.  We use each set of vectors for all displacements. To assess the efficacy of each probing set, we must compare their performance at equal solver cost, which is defined as
\begin{equation}
    C_{p,k} = 12c_{p,k}\sum_{l = 0}^L r_l N_l
    \label{solver_cost_probing}
\end{equation}
where, for each $(p,k)$ probing set, $c_{p,k}$ is the number of colors, and $r_l$ and $N_l$ are the number of solver iterations and number of samples for the level $l$, respectively. By setting a target variance $\epsilon^2$, the number of noise vectors can be calculated using Equation (\ref{num_samples}) for each probing set. Equation (\ref{solver_cost_probing}) can then be used to scale the costs for all sets of probing vectors such that they have equal cost.

\begin{figure}[h!]
\begin{center}
\mbox{\subfigure{\includegraphics[width=0.5\textwidth,scale=0.5]{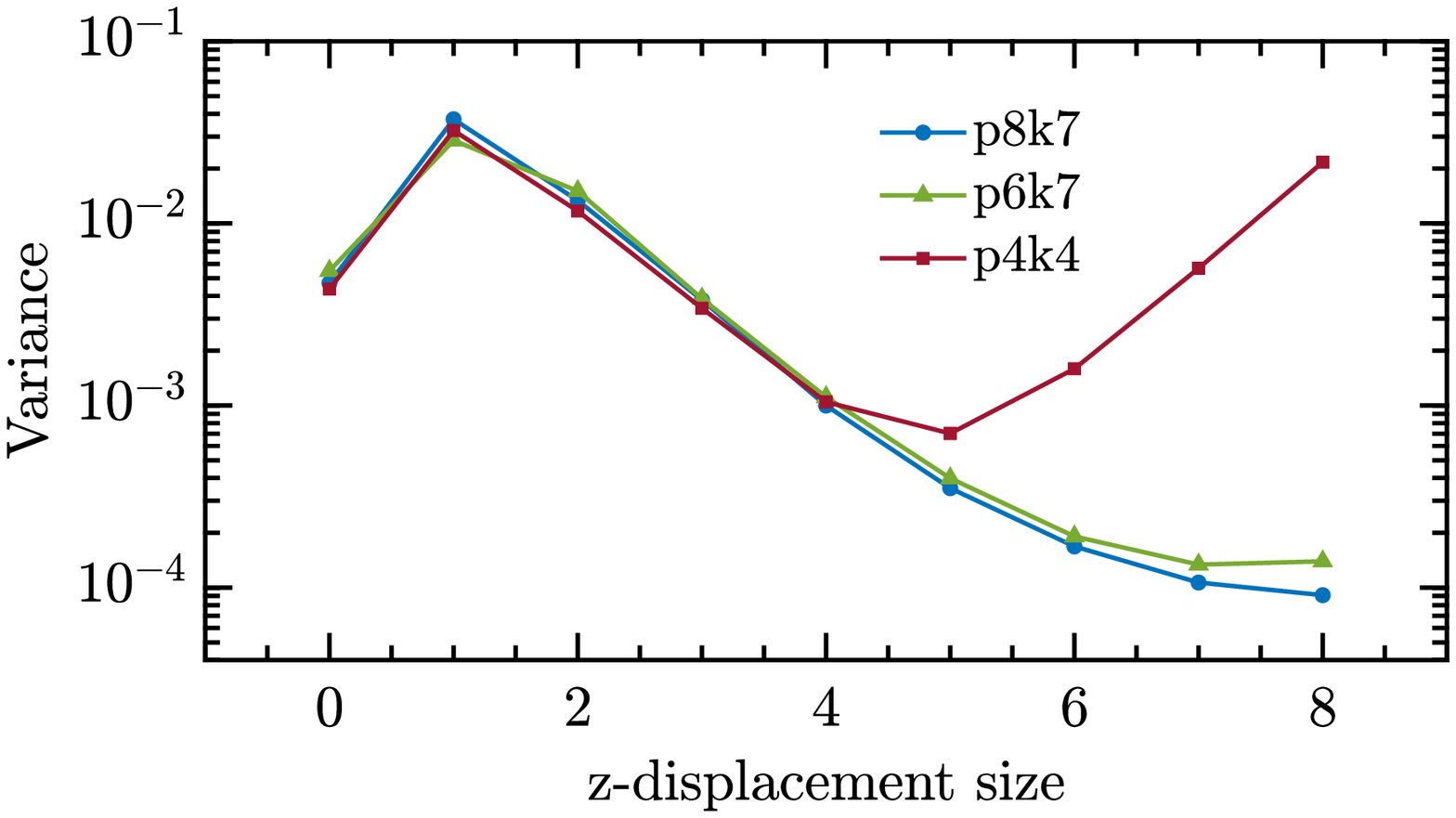}}

\subfigure{\includegraphics[width=0.5\textwidth,scale=0.5]{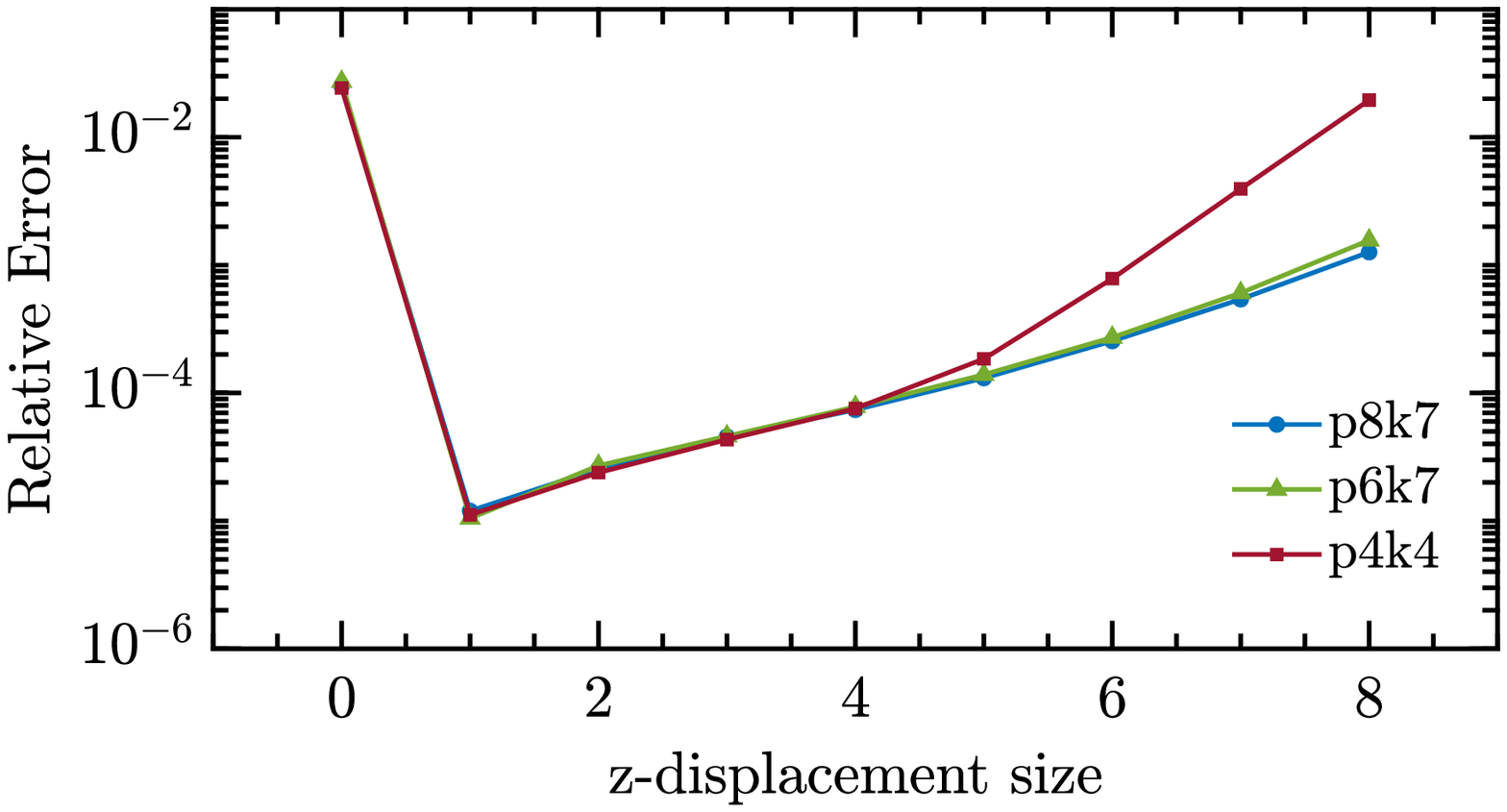} }}
\caption{The FS variance given by Equation (\ref{total_var}) (left) and relative error (right) of $\gamma_3 \Omega_p D^{-1}$ for each set of chosen probing vectors. The optimal shifts were selected as outlined in Sec. \ref{subsec:sample}-\ref{subsec:shift_select}. }
\label{all_prob}
\end{center}
\end{figure}

Figure \ref{all_prob} displays the variance given by Equation (\ref{total_var}) and relative error of estimating the trace of $\gamma_3 \Omega_p D^{-1}$ using the FS method at equal solver cost, where the shifts for each set of probing vectors were chosen through the sampling and interpolation method as outlined in Sec. \ref{subsec:sample}-\ref{subsec:shift_select}. It can be seen that all sets of probing vectors display equal performance for low displacements, however the variance using $p4k4$ probing vectors is several orders of magnitude larger than that of the other two probing schemes at large displacements, and the quality of the relative error degrades accordingly. It can also be observed that $p8k7$ probing vectors offer a slight reduction in variance at larger displacements than $p6k7$ vectors, while having similar effects at lower displacements. It is for this reason that we use $p8k7$ vectors for all displacements for the duration of this work.




\section{Results}
\label{sec:results}

We test the efficacy of our resulting method in three contexts. 
First, we establish that the interpolation scheme predicts the $V_{total}$ and $C_{FS}$, given by Equations (\ref{tosv}) and (\ref{ml_cost_shifts}), to high accuracy by estimating the $V_{total}$ and $C_{FS}$ using the found optimal $\sigma$. Second, we compare the performance of FS using $\sigma$ to multigrid deflation at constant solver cost on a single gauge configuration. Finally, we show that $\sigma$ can be used across multiple configurations with little performance loss, and show that significant speed up is gained over multigrid deflation for each configuration. All calculations were performed with a lattice of size $32^3 \times 64$ using the Clover discretization with even-odd preconditioning and a quark mass of $m_q = -0.2390$, corresponding to the physical light quark mass of the gauge configuration using the Chroma library \cite{EDWARDS2005832} on 8 KNL nodes. For our solver, we use a multigrid preconditioner \cite{mg_proto,QPhiX} for block FGMRES \cite{bfgmres,ELMAN20051037}, with the near null space being calculated once for $\sigma_0 = 0.0$, and subsequently reused for the larger shifts. The variances were sampled using 5 $\mathbb{Z}_4$ noise vectors with full spin-color dilution and $p8k7$ probing vectors. 

\subsection{Accuracy of Interpolation}
\label{subsec:accuracy}


\begin{table}[!h]
\begin{footnotesize}
    \centering
    \begin{tabular}{|c|c|c|c|c|}
    \hline
    \multicolumn{5}{|c|}{$\gamma_3$} \\
    \hline
         p & Pred. $V_{total}$ & Est. $V_{total}$  & Pred. $C_{FS}$ $(\times 10^5)$ & Est. $C_{FS}$ $(\times 10^5)$ \\
         \hline
         0 & 4.9504 & 5.2968 & 0.2921 & 0.3422 \\
         \hline
         1 & 82.1364 & 99.4092 & 1.4293 & 1.7824 \\
         \hline
         2 & 20.8019 & 23.7536 & 0.7521 & 0.8781 \\
         \hline
         3 & 4.4729 & 4.6869 & 0.2371 & 0.2665 \\
         \hline
         4 & 1.1335 & 1.1263 & 0.0680 & 0.0742 \\
         \hline
         5 & 0.3491 & 0.3578 & 0.0215 & 0.0245 \\
         \hline
         6 & 0.1469 & 0.1528 & 0.0084 & 0.0094 \\
         \hline
         7 & 0.0826 & 0.0887 & 0.0041 & 0.0052 \\
         \hline
         8 & 0.0410 & 0.0367 & 0.0030 & 0.0030 \\
         \hline
    \end{tabular}
    \quad
    \begin{tabular}{|c|c|c|c|c|}
    \hline
    \multicolumn{5}{|c|}{$\gamma_5 \gamma_4$} \\
    \hline
         p & Pred. $V_{total}$ & Est. $V_{total}$  & Pred. $C_{FS}$ $(\times 10^4)$ & Est. $C_{FS}$ $(\times 10^4)$ \\
         \hline
         0 & 18.8176 &  21.6828 & 4.8145 & 5.7190 \\
         \hline
         1 & 9.3321 & 9.7361 & 3.9570 & 4.4446 \\
         \hline
         2 & 2.4040 & 2.5149 & 1.4491 & 1.6664 \\
         \hline
         3 & 0.7998 & 0.8279 & 0.4895 & 0.5648 \\
         \hline
         4 & 0.3110 & 0.3111 & 0.1812 & 0.2080 \\
         \hline
         5 & 0.1509 & 0.1443 & 0.0796 & 0.0875 \\
         \hline
         6 & 0.0581 & 0.0464 & 0.0408 & 0.0389 \\
         \hline
         7 & 0.0356 & 0.0279 & 0.0253 & 0.0227 \\
         \hline
         8 & 0.0320 & 0.0234 & 0.0217 & 0.0185 \\
         \hline
    \end{tabular}    
    \caption{The predicted and estimated $V_{total}$  and $C_{FS}$ of $\Gamma \Omega_p D^{-1}$ for the $\Gamma = \gamma_3, \gamma_5 \gamma_4$ operators for each displacement of size $p$. The shifts used for each $(\Gamma,\Omega_p)$ pair are those coming from a minimization of $C_{FS}$ for that $(\Gamma,\Omega_p)$ pair.}
    \label{tab:percent_errors}
    \end{footnotesize}
\end{table}

In order to establish the accuracy of the interpolation scheme, we use the predicted shifts from the interpolation and shift selection scheme with the same gauge configuration to estimate $V_l$ and $V_L$ and therefore calculate $V_{total}$ given by Equation (\ref{tosv}). While many different combinations of $\Gamma$ and $\Omega_p$ may be desired in an experimental setting, and thus many different trace estimations, it is too expensive to verify the accuracy of the interpolation scheme for every combination of $\Gamma$ and $\Omega_p$, as the found optimal shifts are different for each individual combination of $\Gamma$ and $\Omega_p$. We therefore choose to restrict our sampling to variances associated with $\Gamma = \gamma_3$ and $\Gamma = \gamma_5 \gamma_4$, with $\Omega_p$ corresponding to displacements of size $p = \{0,...,8\}$, as these are of particular importance in ongoing calculations pertaining to GDPs.

The estimate of the $V_{total}$ was performed with 5 independent $\mathbb{Z}_4$ noise vectors from those used for sampling and are independent for each level, so we expect the results to be decorrelated. Table \ref{tab:percent_errors} displays the predicted $V_{total}$ and $C_{FS}$ as well as the thei estimated values of $V_{total}$ and $C_{FS}$ for the $\gamma_3$ and $\gamma_5 \gamma_4$ operators. For the purposes of stochastic trace estimation, we observe that our interpolation scheme is able to predict $V_{total}$ to high accuracy. Many of the displacements of size $p \geq 3$ have predicted $V_{total}$ with one digit of accuracy. We also observe that the multilevel cost $C_{FS}$ is well predicted, despite the error introduced through the interpolation of the solver costs. This indicates that the shifts selected are close to those that would give the true minimum cost of the multilevel trace estimation.  

\subsection{Comparison to Multigrid Deflation}
\label{subsec:compare_mgdefl}

We compare the estimated variance of FS using the optimal shifts  to the variance achieved by multigrid deflation with probing and to random noise. In the case of multigrid deflation, we deflate the smallest 400 singular pairs of the coarsened lattice Dirac operator computed with the PRIMME library \cite{primme,svds_software}. However, $p8k7$ probing, which only uses 16 colors, was not as effective when combined with multigrid deflation in reducing the variance over all displacements. We therefore use a probing scheme $p5k8$  that has 256 colors and thus offers more effective variance reduction, making it a more difficult test for FS.

The shifts used are those resulting from the optimization of the ($\Gamma, \Omega_p$) pair ($\gamma_3, \Omega_4$), which gave $\sigma = [0,~10^{-5},~ 0.053,~0.146,~0.360,~0.618,~1.000]$. We use this particular pair for three reasons: 1) the error in the interpolation was found to be much smaller when using $\Omega_p = \Omega_4$. 2) Optimizing the shifts at $\Omega_p = \Omega_4$ provided an additional variance reduction at low displacements compared to optimizing the shifts for high displacements. 3) $\gamma_3$ was found to be more sensitive to the shifts than $\gamma_5 \gamma_4$ at low displacements. When the shifts found from an optimization of $\gamma_5 \gamma_4$ were used, the performance of the trace estimator of $\gamma_3 \Omega_p D^{-1}$ suffered.

To compare the three methods, we first calculate the number of samples required per level for FS to reach a target variance $\epsilon^2 = 0.001$ using Equation (\ref{num_samples}). The total solver cost for FS is given by Equation (\ref{solver_cost_probing}), $C = C_{8,7}$.
The total cost for multigrid deflation is defined for a single level as $c_d r_d N_d$, where $c_d$ and $r_d$ are the number of colors and number of iterations for deflation, respectively. Equating the total solver costs of both methods yields
    \begin{equation}
        N_{d} = \frac{C}{c_{d}r_{d}}
        \label{num_samples_defl}
    \end{equation}
for the required number of samples for multigrid deflation.  The same calculation can be performed for the number of samples required for random noise. Since the variance decreases linearly with the number of samples, we can estimate the variance for all ($\Gamma,\Omega_p$) pairs after $N_l$ steps of Hutchinson on the appropriate level using Equation (\ref{total_var}).

\begin{figure}[!h]
\subfigure{\includegraphics[trim=1 40 10 200,clip=true,width=0.9\textwidth,scale=1.0]{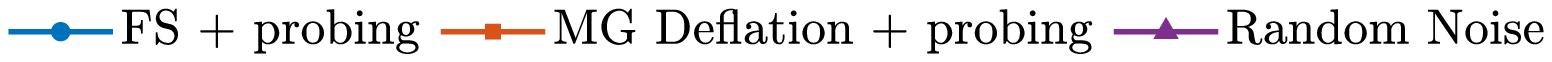}}
\subfigure{\includegraphics[width=1.0\textwidth,scale=0.75]{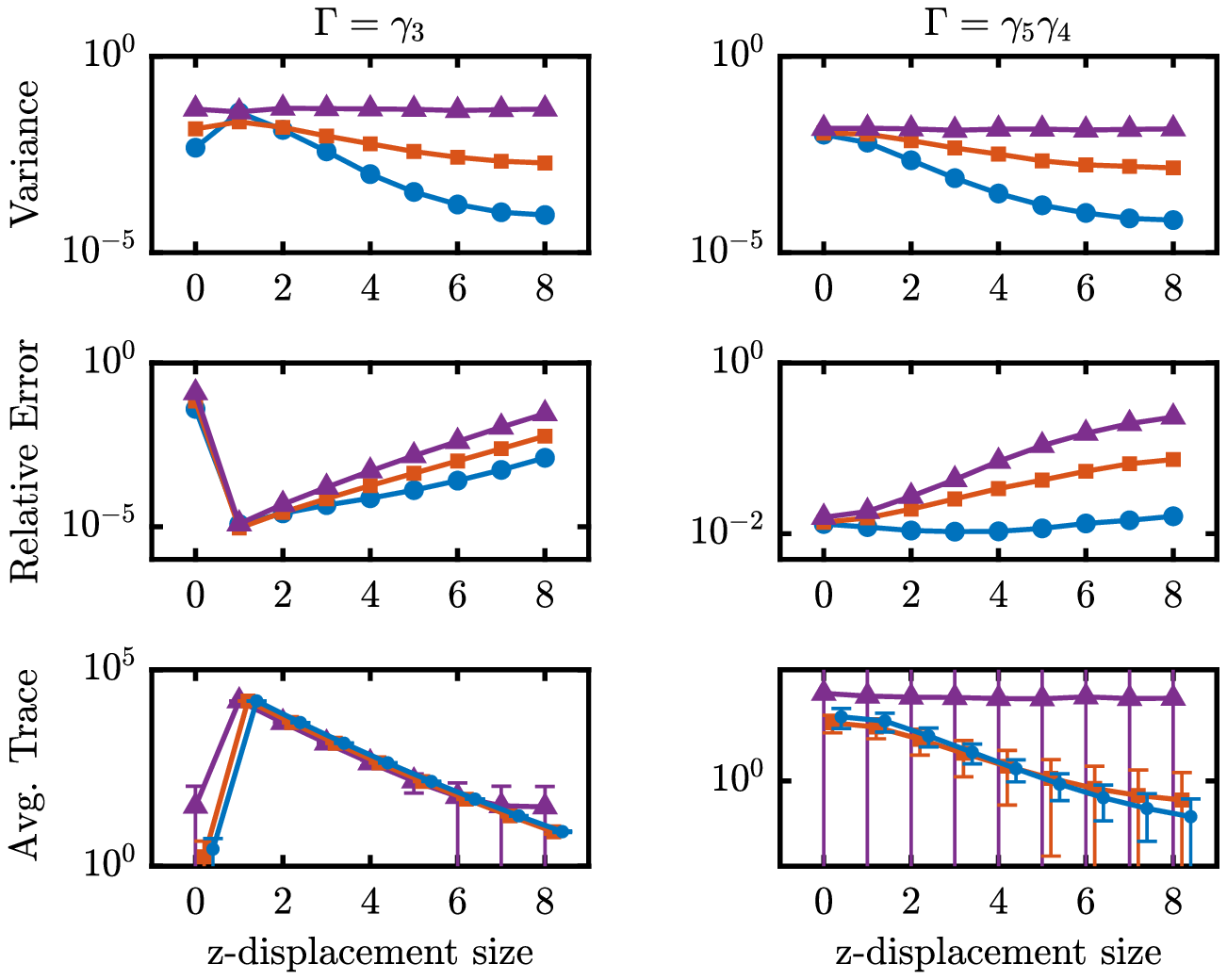}}













\caption{The estimated variance (top), relative error (middle) for $\Gamma = \gamma_3$ (left) and $\Gamma = \gamma_5 \gamma_4$ (right) after $N_l$ samples with p8k7 probing vectors using FS compared to MG deflation with p5k8 probing vectors after $N_d$ samples for a target variance of $\epsilon^2 = 0.001$ at equal solver cost. (Bottom) The average trace for each $\Gamma$ operator using $N_s = 5$ samples. The individual points have been shifted along the $x$-axis for clarity. The shifts used in FS were selected from an optimization of $(\Gamma,\Omega_p)=(\gamma_3,\Omega_4)$.}
\label{vector_axial_vector_single_config}
\end{figure}

Figure \ref{vector_axial_vector_single_config} shows the results of such a calculation for the $\gamma_3$ and $\gamma_5 \gamma_4$ operator as a function of the z displacement. For both the $\gamma_3$ and $\gamma_5 \gamma_4$, we observe that FS is competitive with deflation at low displacements, despite not being optimized for those displacements. As the displacement increases, FS becomes vastly superior to that of deflation. In the case of the $\gamma_3$ operator, FS displays a speedup (the ratio of the variances) of approximately 20 over deflation and approximately 500 over random noise at the largest displacement. In the case of the $\gamma_5 \gamma_4$ operator, FS displays a speedup of approximately 20 over deflation and approximately 200 over random noise at the largest displacement. In the case of the relative error, we also observe improvements for both the $\gamma_3$ and $\gamma_5 \gamma_4$ operators. The average trace reported for each method in Figure \ref{vector_axial_vector_single_config} is computed with $N_s = 5$ noise vectors, with the error bars being the associated standard deviation. For results using other $\Gamma$ operators, see \ref{app:speedup}.

\subsection{Multiple Configurations}
\label{subsec:mult_cfg}

In an actual trace estimation setting, it is too costly to perform the sampling and interpolation method for each configuration, as hundreds or even thousands of configurations may be required, each of them requiring an estimation of the trace of $\Gamma \Omega_p D^{-1}$. Therefore, it is important that the same set of shifts behave similarly well for multiple configurations within a particular gauge ensemble. To this end, we examine the estimated $V_{total}$ of ten configurations using the same set of shifts as in Sec. \ref{subsec:compare_mgdefl}. The estimation is performed with independent noise vectors for each level and each configuration. The gauge configurations are separated by 100 HMC steps. More information regarding these gauge configurations can be found in \cite{PhysRevD.93.114506}. Table \ref{tab:multi_config} displays the mean $V_{total}$ and the relative standard deviation of $V_{total}$ across the ten configurations for all displacements of the $\gamma_3$ and $\gamma_5 \gamma_4$ operators. The relative standard deviation was calculated with jackknife resampling. As can be seen from the relative standard deviation of both operators, $V_{total}$ has very little deviation from one configuration to another. This indicates that the shifts found through the sampling and interpolation of variances from one configuration can be used for others within the same ensemble without impacting performance.  

\begin{table}[!h]
\begin{footnotesize}
    \centering
    \begin{tabular}{|c|c|c|}
    \hline
    \multicolumn{3}{|c|}{$\gamma_3$} \\
    \hline
         Displacement & Mean $V_{total}$ & Rel. Std. Dev. $V_{total}$  \\
         \hline
         0 & 5.4108 &  0.0052 \\
         \hline
         1 & 41.4419 & 0.0029 \\
         \hline
         2 & 16.0299 & 0.0038 \\
         \hline
         3 & 4.7654 & 0.0047 \\
         \hline
         4 & 1.1777 & 0.0056 \\
         \hline
         5 & 0.3883 & 0.0058 \\
         \hline
         6 & 0.1772 & 0.0061 \\
         \hline
         7 & 0.1123 & 0.0083 \\
         \hline
         8 & 0.0932 & 0.0100 \\
         \hline
    \end{tabular}
    \quad
    \begin{tabular}{|c|c|c|}
    \hline
    \multicolumn{3}{|c|}{$\gamma_5 \gamma_4$} \\
    \hline
         Displacement & Mean $V_{total}$ & Rel. Std. Dev. $V_{total}$  \\
         \hline
         0 & 11.3365 &  0.0022 \\
         \hline
         1 & 7.4778 & 0.0024 \\
         \hline
         2 & 2.6171 & 0.0023 \\
         \hline
         3 & 0.8827 & 0.0036 \\
         \hline
         4 & 0.3455 & 0.0065 \\
         \hline
         5 & 0.1727 & 0.0080 \\
         \hline
         6 & 0.1079 & 0.0106 \\
         \hline
         7 & 0.0826 & 0.0114 \\
         \hline
         8 & 0.0733 & 0.0130 \\
         \hline
    \end{tabular}    
    \caption{The mean $V_{total}$ and relative standard deviation of $V_{total}$ for ten configurations from the same gauge ensemble. The relative standard deviation was calculated through jackknife resampling.}
    \label{tab:multi_config}
    \end{footnotesize}
\end{table}

We also examine the performance of FS in comparison to multigrid deflation over multiple configurations for the $(\Gamma,\Omega_p) = (\gamma_3,\Omega_4)$ pair. In order to do so, we estimate $V_{total}$ for each method, and calculate the number of noise vectors required for each method to reach a target variance $\epsilon^2 = 0.001$.
The estimated wallclock time is calculated by using the average wallclock time for the solver to converge for one set of linear equations.
Table \ref{tab:multi_speedup} displays a consistent speedup of the estimated wallclock time of FS over multigrid deflation for each configuration.  
\begin{table}[!h]
    \centering
    \begin{footnotesize}
    \begin{tabular}{|c|c|c|c|c|c|}
    \hline
         Config. \# & 1 & 2 & 3 & 4 & 5 \\
         \hline
         Est. Speedup & 4.8436 & 5.4360 & 4.8494 & 4.5541 & 5.0838 \\
         \hline
         Config. \# & 6 & 7 & 8 & 9 & 10 \\
         \hline
         Est. Speedup & 3.4911 & 4.9955 & 4.5245 & 4.5861 & 5.7280 \\
         \hline
    \end{tabular}
    \caption{The estimated speedup of FS over multigrid deflation for reaching a target variance $\epsilon^2 = 0.001$ for the $(\gamma_3,\Omega_4)$ pair. }
    \label{tab:multi_speedup}
    \end{footnotesize}
\end{table}

\section{Summary}
\label{sec:conc}
We have developed a sampling and interpolation scheme that is able to predict variances to high accuracy and allows us to select a set of near optimal shifts for minimizing the cost of the multilevel trace estimation using the FS method in conjunction with probing for a particular $(\Gamma,\Omega_p)$ pair. We have shown that the use of these near optimal shifts in FS with probing is competitive with or more effective than multigrid deflation with probing, in particular for reducing the variance of the trace estimation corresponding to large lattice displacements. We have also shown that the shifts can be safely reused for multiple configurations with very little impact to overall performance of the trace estimation for each configuration,  displaying consistent speed up over multigrid deflation.

There is also a question of optimization regarding the techniques to use at each level. While we used all variance reduction techniques at our disposal with FS, it is possible that there could be savings by forgoing the use of probing vectors and/or spin-color dilution for certain levels, especially those where the variance is small. A remaining avenue of exploration is an extension of the FS method itself. Since the multilevel analysis allows us to examine the cost of the multilevel trace estimation, it is possible that further recursion of the telescoping series could result in additional gain of performance. We leave such a study for future work. 

\section{Acknowledgements}
\label{sec:ack}
This research was supported by the Exascale Computing Project (ECP), Project Number: 17-SC-20-SC, a collaborative effort of the U.S. Department of Energy, Office of Science and the National Nuclear Security Administration. In addition, KO was supported in part by the 
U.S. Department of Energy, Office of Science, Office of Nuclear Physics Grants DE-FG02-04ER41302, and  DE-AC05-06OR23177.
Computations for this work were carried out in part on facilities of the USQCD Collaboration (D. Richards, HadStruc collaboration allocation at JLab), which are funded by the Office of Science of the U.S. Department of Energy, and  in part using computing facilities at William \& Mary which were provided by contributions from the National Science Foundation (MRI grant PHY-1626177), and the Commonwealth of Virginia Equipment Trust Fund. 

\bibliographystyle{plain}
\bibliography{ref}

\appendix
\section{}
\label{app:speedup}
Table \ref{speedup_table} displays the speedup of FS over multigrid deflation for all $(\Gamma,\Omega_p)$ pairs computed. Since FS and multigrid deflation are examined at equal cost, we define the speed up to be the ratio of the variances:
\begin{center}
    \begin{equation}
        Speedup = \frac{V_{FS}}{V_{Defl}}.
        \label{app_speedup}
    \end{equation}
\end{center}

\begin{table}[!t]
\centering
\begin{small}
    \begin{tabular}{|c|c|c|c|c|c|c|c|c|c|}
    \hline
    \multicolumn{10}{|c|}{\textbf{Speedup}} \\
    \hline
    $\Gamma$ & \multicolumn{9}{|c|}{Displacement} \\
    \hline
    & 0 & 1 & 2 & 3 & 4$^*$ & 5 & 6 & 7 & 8  \\
    \hline
$I$ & 0.13 & 0.16 & 0.25 & 0.53 & 1.36 & 3.13 & 5.89 & 8.58 & 10.24  \\
\hline
$\gamma_1$ & 0.40 & 0.92 & 2.12 & 4.81 & 8.95 & 12.72 & 15.78 & 17.28 & 18.02  \\
\hline
$\gamma_2$ & 0.39 & 0.91 & 2.10 & 4.90 & 9.19 & 13.47 & 15.95 & 17.18 & 18.43  \\
\hline
$\gamma_1 \gamma_2$ & 0.66 & 1.08 & 1.48 & 2.52 & 4.34 & 6.72 & 9.16 & 10.99 & 11.95  \\
\hline
$\gamma_3^*$ & 3.00 & 0.57 & 1.16 & 2.40 & 5.36 & 10.01 & 14.44 & 16.95 & 18.62  \\
\hline
$\gamma_1 \gamma_3$ & 0.45 & 0.68 & 1.17 & 2.33 & 4.27 & 6.79 & 9.32 & 11.24 & 12.45  \\
\hline
$\gamma_2 \gamma_3$ & 0.48 & 0.69 & 1.17 & 2.29 & 4.33 & 7.03 & 9.74 & 11.60 & 12.45  \\
\hline
$\gamma_5 \gamma_4$ & 1.07 & 1.54 & 3.03 & 5.68 & 9.45 & 12.85 & 15.33 & 16.65 & 17.80  \\
\hline
$\gamma_4$ & 0.55 & 1.07 & 2.23 & 4.84 & 9.13 & 13.25 & 16.45 & 18.00 & 18.48  \\
\hline
$\gamma_1 \gamma_4$ & 0.40 & 0.62 & 1.09 & 2.15 & 4.09 & 6.69 & 9.34 & 11.57 & 12.73  \\
\hline
$\gamma_2 \gamma_4$ & 0.43 & 0.64 & 1.10 & 2.11 & 4.03 & 6.58 & 9.21 & 11.28 & 12.38  \\
\hline
$\gamma_3 \gamma_5$ & 0.98 & 1.83 & 3.57 & 6.19 & 9.29 & 12.45 & 14.33 & 16.16 & 16.98  \\
\hline
$\gamma_3 \gamma_4$ & 0.64 & 0.62 & 0.95 & 1.94 & 3.79 & 6.36 & 9.10 & 11.28 & 12.53  \\
\hline
$\gamma_2 \gamma_5$  & 1.23 & 1.71 & 3.13 & 5.48 & 8.84 & 12.15 & 14.41 & 16.31 & 17.84  \\
\hline
$\gamma_1 \gamma_5$ & 1.22 & 1.71 & 3.13 & 5.60 & 8.87 & 12.40 & 15.00 & 16.28 & 17.31  \\
\hline
$\gamma_5$ & 0.17 & 0.25 & 0.50 & 1.10 & 2.51 & 5.09 & 8.58 & 11.22 & 12.72  \\
\hline

    \end{tabular}
    \caption{The speed up of FS over multigrid deflation for all tested combinations of $\Gamma,\Omega_p$. $^*$ denotes the $(\Gamma,\Omega_p)$ pair used for optimization.}
    \label{speedup_table}
    \end{small}
\end{table}
We remark that the performance of FS for many of the operators listed in Table \ref{speedup_table} at low displacements is worse than that of deflation. In an actual physics simulation, this is not much of a concern since the relative error is much larger for traces that are non vanishing under large statistics (many gauge configurations) at low displacements. For traces that are vanishing under large statics, the trace will average towards zero. We also observe from table \ref{speedup_table} that the speedup increases as the lattice displacement increases, resulting in an order of magnitude improvement at a displacement of 8. 
\end{document}